\documentclass[11pt]{article}
\usepackage{jheppub}
\usepackage{epsfig}
\usepackage{amssymb}
\usepackage{amsmath}
\usepackage{graphicx, subfigure, array, placeins, float}

\usepackage{blkarray}

\usepackage{tcolorbox}
\tcbuselibrary{fitting}
\usepackage{makecell}
\usepackage{multirow}
\usepackage{epstopdf}
\usepackage{extarrows}
\usepackage{dsfont,bbm}
\usepackage{shuffle}
\usepackage{blkarray}

\usepackage{amsmath}
\graphicspath{{./figs/}}

\usepackage{subcaption} 

\newcommand{\ap}{{\rm a}}
\newcommand{\am}{\textrm{-a}}
\newcommand{\bp}{{\rm b}}
\newcommand{\bm}{\textrm{-b}}
\newcommand{\cp}{{\rm c}}
\newcommand{\cm}{\textrm{-c}}
\newcommand{\dpp}{{\rm d}}
\newcommand{\dmm}{\textrm{-d}}

\usepackage{tikz}
\usepackage{adjustbox}
\usetikzlibrary{decorations.markings}
\usetikzlibrary{arrows.meta}
\tikzset{
    midarrow/.style={
        decoration={markings,
            mark=at position 0.58 with {\arrow{stealth}}
        },
        postaction={decorate}
    }
}
\tikzset{
    midreversearrow/.style={
        decoration={markings,
            mark=at position 0.42 with {\arrowreversed{stealth}}
        },
        postaction={decorate}
    }
}
\newcommand{\WLb}[1]{
\adjustbox{valign=c}{\begin{tikzpicture}[scale=0.5]
	\draw [#1] (0, 0)--(1, 0);
	\draw (1, 0)--(2, 0)--(2, 1)--(1, 1)--(0, 1)--(0, 0);
\end{tikzpicture}}}
\newcommand{\WLc}[1]{
\adjustbox{valign=c}{\begin{tikzpicture}[scale=0.5]
	\draw [#1] (0, 0)--(1, 0);
	\draw (1, 0)--(1.2, 1);
	\draw [fill=white, draw=white] (1.1,0.5) circle (0.2);
	\draw (1.2, 1)--(2.2, 1)--(2.2, 0)--(1.2, 0)--(1, 1)--(0, 1)--(0, 0);
\end{tikzpicture}}}
\newcommand{\WLj}[1]{
\adjustbox{valign=c}{\begin{tikzpicture}[scale=0.5]
	\draw [#1] (0, 0)--(1, 0);
	\draw (1, 0)--(2, 0)--(2, -1)--(1, -1);
	\draw [fill=white, draw=white] (1, 0) circle (0.1);
	\draw (1, -1)--(1, 0)--(1, 1)--(0, 1)--(0, 0);
\end{tikzpicture}}}
\newcommand{\WLk}[1]{
\adjustbox{valign=c}{\begin{tikzpicture}[scale=0.5]
	\draw [#1] (0, 0)--(0.8, 0);
	\draw (0.8, 0)--(0.8, -1)--(1.8, -1)--(1.8, 0)--(1, 0)--(1, 1)--(0, 1)--(0, 0);
\end{tikzpicture}}}
\newcommand{\WLbr}{
\adjustbox{valign=c}{\begin{tikzpicture}[scale=0.5]
	\draw [red,midarrow] (1, 0)--(2, 0);
	\node[scale=0.8] at (1.5, -0.3) {$\epsilon$};
	\draw (2, 0)--(2, 0.2)--(1, 0.2)--(1, 1)--(0, 1)--(0, 0)--(1,0);
\end{tikzpicture}}}
\newcommand{\WLdaa}[2]{
\adjustbox{valign=c}{\begin{tikzpicture}[scale=0.5]
	\draw [#1] (0, 0)--(1, 0);
	\draw (1, 0)--(1, 1)--(0, 1)--(0, 0);
	\draw [#2] (1.2, 0)--(2.2, 0);
	\draw (2.2, 0)--(2.2, 1)--(1.2, 1)--(1.2, 0);
\end{tikzpicture}}}

\newcommand{\WLdac}[2]{
\adjustbox{valign=c}{\begin{tikzpicture}[scale=0.5]
	\draw [#1] (0, 0)--(1, 0);
	\draw (1, 0)--(1, 1)--(0, 1)--(0, 0);
	\draw [#2] (1.2, 0)--(2.2, 0);
	\draw (2.2, 0)--(2.2, -1)--(1.2, -1)--(1.2, 0);
\end{tikzpicture}}}

\newcommand{\wle}{
\adjustbox{valign=c}{\begin{tikzpicture}[scale=0.5]
	\draw (0, -1)--(0, 1)--(-1, 1)--(-1, 0)--(-1.2, -1);
    \draw [fill=white, draw=white] (-1.1,-0.5) circle (0.2);
	\draw (-1.2, -1)--(-3, -1)--(-3, 0)--(-1.2, 0)--(-1, -1)--(0, -1);
\end{tikzpicture}}}

\newcommand{\wlf}{
\adjustbox{valign=c}{\begin{tikzpicture}[scale=0.5, blue]
	\draw (0, -0.1)--(0, -1)--(-1, -1)--(-1, -0.1)--(0, 0.1)--(0, 1)--(-1.2, 1)--(-1.2, -1);
	\draw [fill=white, draw=white] (-1.2,0) circle (0.2);
	\draw [fill=white, draw=white] (-0.5,0) circle (0.2);
	\draw (-1.2, -1)--(-3, -1)--(-3, 0)--(-1, 0)--(0, -0.1);
\end{tikzpicture}}}

\newcommand{\wlk}{
\adjustbox{valign=c}{\begin{tikzpicture}[scale=0.5]
	\draw  (0, 0)--(0, 1)--(-2, 1)--(-2, 0)--(0, 0);
\end{tikzpicture}}}

\newcommand{\wll}{
\adjustbox{valign=c}{\begin{tikzpicture}[scale=0.5]
	\draw (0, 0)--(0, 1);
 	\draw (0, 1)--(-1, 1)--(-1, -1);
	\draw [fill=white, draw=white] (-1,0) circle (0.2);
	\draw (-1, -1)--(-3, -1)--(-3, 0)--(0, 0);
\end{tikzpicture}}}

\newcommand{\wlm}{
\adjustbox{valign=c}{\begin{tikzpicture}[scale=0.5, blue]
	\draw (0,0)--(0, 2)--(1, 2)--(1, 1.2);
	\draw [fill=white, draw=white] (0.25,2) circle (0.2);
	\draw (1, 1.2)--(0.2, 1.2)--(0.2, 2.2)--(1.2, 2.2);
	\draw [fill=white, draw=white] (0,1) circle (0.2);
	\draw (1.2, 2.2)--(1.2, 1)--(-2, 1)--(-2, 0)--(0, 0);
\end{tikzpicture}}}

\newcommand{\wln}{
\adjustbox{valign=c}{\begin{tikzpicture}[scale=0.5]
	\draw  (0, 0)--(0, 1.2)--(-1, 1.2)--(-1, 2.2)--(1, 2.2)--(1, 1);
	\draw [fill=white, draw=white] (0,0.95) circle (0.2);
	\draw (1, 1)--(-2, 1)--(-2, 0)--(0, 0);
\end{tikzpicture}}}

\newcommand{\wlo}{
\adjustbox{valign=c}{\begin{tikzpicture}[scale=0.5, blue]
	\draw (0.1, 1.1)--(-0.1, 2)--(-1, 2)--(-1, 1.1);
	\draw [fill=white, draw=white] (0,1.5) circle (0.2);
	\draw (-1, 1.1)--(-0.1, 1.1)--(0.1, 2)--(1, 2)--(1, 1)--(1, 1)--(-2, 1);
	\draw [fill=white, draw=white] (0.1,0.85) circle (0.2);
	\draw (-2, 1)--(-2, 0)--(0.1, 0)--(0.1, 1.1);
\end{tikzpicture}}}

\newcommand{\wlp}{
\adjustbox{valign=c}{\begin{tikzpicture}[scale=0.5]
	\draw  (0.1, 0)--(1, 0)--(1, 1)--(-0.1, 1)--(-0.1, 0)--(-2, 0)--(-2, -1)--(0.1, -1)--(0.1, 0);
\end{tikzpicture}}}

\newcommand{\wlq}{
\adjustbox{valign=c}{\begin{tikzpicture}[scale=0.5]
	\draw (0, 0)--(0, 1);
 	\draw (0, 1)--(-0.9, 1)--(-0.9, -1);
	\draw [fill=white, draw=white] (-0.9,0) circle (0.2);
	\draw (-0.9, -1)--(-3, -1)--(-3, 0)--(-2, 0)--(-2, 1)--(-1.1, 1)--(-1.1, 0)--(0, 0);
\end{tikzpicture}}}

\newcommand{\wlr}{
\adjustbox{valign=c}{\begin{tikzpicture}[scale=0.5, blue]
	\draw (1,1.1)--(-1, 1.1);	
	\draw [fill=white, draw=white] (0.1,1.1) circle (0.2);
	\draw [fill=white, draw=white] (-0.1,1.1) circle (0.2);
	\draw (-1, 1.1)--(-1, 2)--(-0.1, 2)--(-0.1, 0.9)--(-2, 0.9)--(-2, 0)--(0.1, 0)--(0.1, 2)--(1, 2)--(1,1.1);
\end{tikzpicture}}}

\newcommand{\wlt}{
\adjustbox{valign=c}{\begin{tikzpicture}[scale=0.5]
	\draw (0,0.1)--(0,1)--(-1, 1)--(-1, 0.1)--(0, -0.1);
	\draw [fill=white, draw=white] (-0.5,0) circle (0.2);
	\draw  (0, -0.1)--(0, -1)--(-3, -1)--(-3, -0.1)--(-1, -0.1)--(0, 0.1);
\end{tikzpicture}}}
\newcommand{\wlu}{
\adjustbox{valign=c}{\begin{tikzpicture}[scale=0.5,blue]
	\draw (0.1,0)--(1, 0)--(1, -0.9)--(-2, -0.9)--(-2, -2)--(-0.1, -2)--(-0.1,-1.1)--(1,-1.1)--(1,-2)--(0.1, -2);
	\draw [fill=white, draw=white] (0.15,-1) circle (0.2);
	\draw (0.1, -2)--(0.1, 0);
\end{tikzpicture}}}

\newcommand{\wlaa}{
\adjustbox{valign=c}{\begin{tikzpicture}[scale=0.5]
	\draw (0,0)--(1,0)--(1,1)--(2,1)--(2,2)--(0,2)--(0,0);
\end{tikzpicture}}}

\newcommand{\wlbb}{
\adjustbox{valign=c}{\begin{tikzpicture}[scale=0.5]
	\draw (-0.1, 0)--(-1,0)--(-1,-2)--(-0.1,-2)--(-0.1,-1)--(0.1,0)--(1,0)--(1,-1)--(0.1,-1);
	\draw [fill=white, draw=white] (0,-0.5) circle (0.2);
	\draw (0.1,-1)--(-0.1,0);
\end{tikzpicture}}}

\newcommand{\wlcc}{
\adjustbox{valign=c}{\begin{tikzpicture}[scale=0.5]
	\draw (-0.1, 0)--(-1,0)--(-1,-2)--(-0.1,-2)--(-0.1,-1)--(0.1,0)--(1,0)--(1,-2)--(0.1,-2);
	\draw [fill=white, draw=white] (0,-0.5) circle (0.2);
	\draw (0.1,-2)--(0.1,-1)--(-0.1,0);
\end{tikzpicture}}}

\newcommand{\wldd}{
\adjustbox{valign=c}{\begin{tikzpicture}[scale=0.5,blue]
	\draw (-1,0)--(1,0)--(1,1)--(0,1);
	\draw [fill=white, draw=white] (-0.1,0) circle (0.25);
	\draw (0,1)--(0,-1)--(-1.2,-1)--(-1.2,1)--(-0.2,1)--(-0.2, 0)--(-0.2,-0.8)--(-1,-0.8)--(-1,0);
\end{tikzpicture}}}

\newcommand{\wlee}{
\adjustbox{valign=c}{\begin{tikzpicture}[scale=0.5,blue]
	\draw (-0.2, 0)--(1,0)--(1,1)--(0,1);
	\draw [fill=white, draw=white] (0.05,0) circle (0.2);
	\draw (0,1)--(0,-1)--(-1.3,-1)--(-1.3,1)--(-0.3,1)--(-0.32, 0)--(-1.1,0)--(-1.1,-0.8)--(-0.2,-0.8)--(-0.2, 0);
\end{tikzpicture}}}

\newcommand{\wlff}{
\adjustbox{valign=c}{\begin{tikzpicture}[scale=0.5,blue]
	\draw (-0.2,-0.2)--(1.1,-0.2);
	\draw [fill=white, draw=white] (0.1,-0.2) circle (0.2);
	\draw (1.1,-0.2)--(1.1,-1)--(0.2,-1)--(0.2,0)--(1.1,0)--(1.1,1)--(0,1)--(0,-1)--(-1.2,-1)--(-1.2,1)--(-0.2,1)--(-0.2,-0.2);
\end{tikzpicture}}}

\newcommand{\wlgg}{
\adjustbox{valign=c}{\begin{tikzpicture}[scale=0.5]
	\draw (0,0)--(1,0)--(1,1)--(0,1)--(0,0);
\end{tikzpicture}}}

\newcommand{\wlhh}{
\adjustbox{valign=c}{\begin{tikzpicture}[scale=0.5]
	\draw (-0.2,-0.8)--(-0.2,1)--(-1.2,1);
	\draw [fill=white, draw=white] (-0.1,0) circle (0.2);
	\draw (-1.2,1)--(-1.2,-1)--(0,-1)--(0,0);
	\draw (0,0)--(-1,0)--(-1,-0.8)--(-0.2,-0.8);
\end{tikzpicture}}}

\newcommand{\wlii}{
\adjustbox{valign=c}{\begin{tikzpicture}[scale=0.5]
	\draw (-0.2,-0.8)--(-1,-0.8)--(-1,0.8)--(0,0.8)--(0,-1)--(-1.2,-1)--(-1.2,1)--(-0.2,1);
	\draw [fill=white, draw=white] (-0.25,0.7) circle (0.2);
	\draw (-0.2,1)--(-0.2,-0.8);
\end{tikzpicture}}}

\newcommand{\wljj}{
\adjustbox{valign=c}{\begin{tikzpicture}[scale=0.5,blue]
	\draw (-0.2,-0.8)--(-0.2,1)--(-1.2,1);
	\draw [fill=white, draw=white] (-0.1,0) circle (0.2);
	\draw (-1.2,1)--(-1.2,-1)--(0.2,-1)--(0.2,0)--(1,0)--(1,1)--(0.2,1)--(0,1)--(0,0)--(-1,0)--(-1,-0.8)--(-0.2,-0.8);
\end{tikzpicture}}}

\newcommand{\wlkk}{
\adjustbox{valign=c}{\begin{tikzpicture}[scale=0.5,blue]
	\draw (-0.1,-0.05)--(1,-0.05)--(1,0.9)--(0.1,1.1);
	\draw [fill=white, draw=white] (0.55,1) circle (0.2);
	\draw (0.1,1.1)--(0.1,1.9)--(1,1.9)--(1,1.1)--(0.1,0.9)--(0.1,0.2);
	\draw [fill=white, draw=white] (1,0.25) circle (0.2);
	\draw (0.1,0.2)--(1.2,0.2)--(1.2,2.1)--(-0.1,2.1)--(-0.1,-0.05);
\end{tikzpicture}}}

\newcommand{\wlll}{
\adjustbox{valign=c}{\begin{tikzpicture}[scale=0.5,blue]
	\draw (0.4, 0)--(1.2,0)--(1.2,1)--(0.2,1)--(0.2,-0.8)--(-0.8,-0.8)--(-0.8,0)--(0,0)--(0,1)--(-1,1)--(-1,-1)--(0.4,-1)--(0.4,0);
\end{tikzpicture}}}

\newcommand{\wlmm}{
\adjustbox{valign=c}{\begin{tikzpicture}[scale=0.5,blue]
	\draw (0.4,0)--(0.4, 0.2)--(-0.8,0.2)--(-0.8,1.2)--(0.2,1.2);
	\draw [fill=white, draw=white] (0.1,0.2) circle (0.2);
	\draw [fill=white, draw=white] (-0.8,0.95) circle (0.2);
	\draw (0.2,1.2)--(0.2,0.2)--(0.2,0);
	\draw (0.2,0)--(0.2,-0.8)--(-0.8,-0.8)--(-0.8,0)--(0,0)--(0,1)--(-1,1)--(-1,-1)--(0.4,-1)--(0.4,0);
\end{tikzpicture}}}

\newcommand{\wloo}{
\adjustbox{valign=c}{\begin{tikzpicture}[scale=0.5,blue]
	\draw (1.2,-1.1)--(0.2,-1.1)--(0.2,1);
	\draw [fill=white, draw=white] (0.2,-0.1) circle (0.2);
	\draw (0.2,1)--(1.2,1)--(1.2,0)--(-0.2,0)--(-0.2,-1.1)--(-1.1,-1.1)--(-1.1,1)--(0,1);
	\draw [fill=white, draw=white] (0,0) circle (0.1);
	\draw (0,1)--(0,-0.2)--(1.2,-0.2)--(1.2,-1.1);
\end{tikzpicture}}}

\newcommand{\wlpp}{
\adjustbox{valign=c}{\begin{tikzpicture}[scale=0.5]
	\draw (0,0)--(0,2)--(2,2)--(2,0)--(0,0);
\end{tikzpicture}}}

\newcommand{\wlqq}{
\adjustbox{valign=c}{\begin{tikzpicture}[scale=0.5]
	\draw (0,0)--(0.9,0)--(0.9,0.9)--(2,0.9);
	\draw [fill=white, draw=white] (1.15,0.9) circle (0.2);
	\draw (2,0.9)--(2,0)--(1.1,0)--(1.1,1.1)--(2,1.1)--(2,2)--(0,2)--(0,0);
\end{tikzpicture}}}

\newcommand{\wlrr}{
\adjustbox{valign=c}{\begin{tikzpicture}[scale=0.5]
	\draw [] (0.2, 0)--(1, 0);
	\draw (1, 0)--(1, 1);
	\draw [fill=white, draw=white] (1, 0.8) circle (0.1);
	\draw (1, 1)--(0, 1)--(0, -0.2)--(1.2, -0.2)--(1.2, 0.8)--(0.2, 0.8)--(0.2, 0);
\end{tikzpicture}}}

\newcommand{\wlss}{
\adjustbox{valign=c}{\begin{tikzpicture}[scale=0.5]
	\draw (0, 0)--(1, 0);
	\draw (1, 0)--(1.2, 1);
	\draw [fill=white, draw=white] (1.1,0.5) circle (0.2);
	\draw (1.2, 1)--(2.2, 1)--(2.2, 0)--(1.2, 0)--(1, 1)--(0, 1)--(0, 0);
\end{tikzpicture}}}

\newcommand{\wltt}{
\adjustbox{valign=c}{\begin{tikzpicture}[scale=0.5]
	\draw (0.1,0.2)--(1.1,0.2);
	\draw [fill=white, draw=white] (0.85,0.2) circle (0.2);
	\draw (1.1,0.2)--(1.1,1)--(2,1)--(2,2)--(-0.1,2)--(-0.1,0)--(0.9,0)--(0.9,1)--(0.1,1)--(0.1,0.2);
\end{tikzpicture}}}

\preprint{USTC-ICTS/PCFT-26-02}
\title{Direct and Indirect Loop Equations in Lattice Yang-Mills Theory}
\author[a,b,c]{Xizhe Liu,}
\emailAdd{liuxizhe24@mails.ucas.ac.cn}
\affiliation[a]{School of Fundamental Physics and Mathematical Sciences, Hangzhou Institute for Advanced Study, UCAS, Hangzhou 310024, China}
\affiliation[b]{School of Physical Sciences, University of Chinese Academy of Sciences, Beijing 100049, China}
\affiliation[c]{Institute of Theoretical Physics, Chinese Academy of Sciences, Beijing 100190, China}
\author[a,b,c,d]{Gang Yang}
\emailAdd{yangg@itp.ac.cn}
\affiliation[d]{Peng Huanwu Center for Fundamental Theory, Hefei, Anhui 230026, China}

\abstract{
The dynamics of Wilson loops are governed by an infinite set of Schwinger-Dyson equations and trace relations. In the context of the lattice positivity bootstrap, a central challenge is determining a dynamically independent basis of these operators within a truncated space. We present a systematic framework to address this problem, utilizing a geometric plaquette-cut and subloop-cut strategy to efficiently generate all (local) direct equations. Furthermore, we identify and analyze ``indirect equations", which arise from the elimination of higher-length intermediate loops. We elucidate the origin of these subtle relations and propose a vertex-filtering strategy to construct them. Applying the above framework to SU(2) lattice Yang-Mills theory, we provide explicit counts of independent canonical loops and equations in 2, 3, and 4 dimensions, along with a statistical analysis of their asymptotic growth.
}

\begin{document}

\maketitle
\setcounter{footnote}{0}

\section{Introduction}
\label{sec:introduction}

The bootstrap strategy, which unifies positivity constraints with dynamical equations, has established itself as a potent nonperturbative tool in quantum field theory.
Its application to lattice gauge theories, initiated by Anderson and Kruczenski \cite{Anderson:2016rcw} and significantly advanced in recent works \cite{Kazakov:2022xuh, Kazakov:2024ool, Li:2024wrd, Guo:2025fii}, promises a rigorous, first-principles alternative to Monte Carlo simulations. 
The lattice positivity bootstrap leverages Schwinger-Dyson (SD) loop equations \cite{Makeenko:1979pb,Migdal:1983qrz} as algebraic constraints on Wilson loop expectation values, which are simultaneously bounded by the convex geometry of the positivity matrices.

A critical challenge in the bootstrap program, however, is to efficiently construct the complete set of loop equations. 
The major difficulty lies in the fact that the Wilson loop operators form an infinitely coupled system.
In any practical computation, one must truncate the infinite set of Wilson loops to a finite subset containing operators up to a certain length. The SD equations for an operator in the subset generically produce new operators outside of the subset---a phenomenon we term ``operator leakage". 
This creates an open system of equations with more variables than constraints, making the identification of the full independent algebraic relations within the truncated subspace highly non-trivial. 
While it is well known that in 2D lattice Yang-Mills theory, all loops can be reduced to single-plaquette variables \cite{Migdal:1975zg, Drouffe:1978dn, Gross:1980he, Wadia:2012fr}, this property does not generalize to higher dimensions. A framework capable of efficiently constructing loop equations in general dimensions, without relying on 2D simplifications, remains to be developed.

In this paper, we address this problem in a systematic way. By exploring the structure of loop equations, we classify them into two distinct categories: direct and indirect equations.
We develop a geometric algorithm---based on plaquette and subloop cuts---that constructs the full direct set of SD and trace-reduction equations.
We further identify and analyze the more subtle ``indirect equations"---constraints that emerge only after eliminating higher-length auxiliary loops---and propose a vertex-filtering method to detect them. 
Explicit results are presented for SU(2) Yang-Mills theories across 2 to 4 dimensions, 
and the scaling laws governing the growth of loop operators and equations are also briefly discussed.

Readers familiar with Self-Avoiding Walks (SAW)  (see e.g.~\cite{madras2012self}) may find some similarity at least for part of this study. However, we would like to emphasize some fundamental differences. Unlike counting walks, here we count loop orbits modulo lattice and cyclic symmetry.  More importantly, we consider dynamical equations between loop variables, which go beyond pure geometric studies. In other words, we are not only asking ``How many shape-independent loops are there?" but also asking ``How many dynamics-independent variables are there?"
Our goal is to find the independent loop basis by taking into account their dynamical relations.

The remainder of the paper is organized as follows. 
In Section~\ref{sec:setup}, we give a review of Wilson loops and loop equations. 
Section~\ref{sec:directeqn} discusses the construction of direct equations.
Section~\ref{sec:growthofloopandeqn} presents the growth of canonical loops and direct equations, and discusses a fit of the data.
In Section~\ref{sec:indirecteqn} we focus on the indirect equations.
A discussion is given in Section~\ref{sec:summary}.
A consideration of plane-type loops is given in Appendix~\ref{app:planetype}.
Further details of indirect equations are provided in Appendix~\ref{app:indirecteq}.

\section{Setup of loops and equations}
\label{sec:setup}

In this section, we briefly review the Wilson loop operators in lattice YM theory. We explain the canonical form of the loops and review the SD equations and trace relations. This section is intended to serve as both a review and a setup for the conventions.

\subsection{Wilson loops}

\paragraph{Lattice theory.}

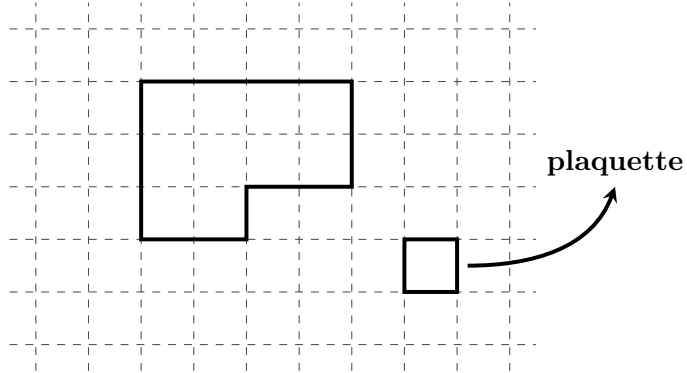
\begin{figure}[t]
    \centering
    \begin{tikzpicture}[scale=0.7]
        \draw[dashed, thin, black!70] (-2.5, -2.5) grid (7.5, 4.5);
%
        \draw[line width=1.5pt] (0, 0) -- (2, 0) -- (2, 1) -- (4, 1) -- (4, 3) -- (0, 3) -- cycle;
%
        \draw[line width=1.5pt] (5, -1) rectangle (6, 0);
%
        \draw[->, line width=1.5pt, >=stealth]
            (6.2, -0.5) to [out=0, in=-110] (9., 1.)
            node[above, font=\bfseries] {plaquette};
    \end{tikzpicture}
    \caption{2D lattice showing a large Wilson loop and a single plaquette.}
    \label{fig:wilson_with_plaquette}
\end{figure}

We consider pure YM theory defined on a hypercubic lattice in $D$ Euclidean dimensions ($D=2,3,4$).
The dynamics are governed by the standard Wilson action \cite{Wilson:1974sk}:
\begin{equation}
S = - {N \over 2\lambda} \sum_P {\rm tr} (U_P + U_P^\dagger) \,,
\end{equation}
where the summation runs over all plaquettes (the elementary square loops) on the lattice. 
The link variables $U_\mu(x)$ are elements of the gauge group, residing on the edges connecting $x$ and $x+\hat\mu$.
The plaquette variable is defined as the ordered product of links around a unit square:
\begin{equation}
	{\rm tr}(U_P) = {\rm tr} \big[ U_\mu(x) U_\nu(x+\hat\mu) U_\mu^\dagger(x+\hat\nu) U_\nu^\dagger(x) \big] =
	\adjustbox{valign=c}{\begin{tikzpicture}[scale=.9]
		\node at (0.5,-0.3) {$U_\mu$};
		\draw [midarrow] (0, 0)--(1, 0);
		\node at (1.3,0.47) {$U_\nu$};
		\draw [midarrow] (1, 0)--(1, 1);
		\node at (0.5,1.3) {$U_\mu^\dagger$};
		\draw [midarrow] (1, 1)--(0, 1);
		\node at (-0.35,0.57) {$U_\nu^\dagger$};
		\draw [midarrow] (0, 1)--(0, 0);
	\end{tikzpicture} 
	} \,,
\end{equation}
where the $U_\mu$'s are fundamental link variables.
A visualization of a 2D lattice configuration containing a single plaquette and a larger loop is shown in Figure~\ref{fig:wilson_with_plaquette}.

The fundamental gauge-invariant observables are the Wilson loops, defined as the trace of the path-ordered product of link variables along a closed curve.
We define the \emph{length} of a Wilson loop as the number of constituent link variables; for instance, the plaquette has length 4 and the larger loop in Figure~\ref{fig:wilson_with_plaquette} is of length 14.

The vacuum expectation value for a Wilson loop operator $W$ is given by
\begin{equation}
\langle W \rangle = Z^{-1} \int {\cal D} U e^{- S} \, W \,, \quad 
Z = \int {\cal D} U e^{- S} = \int \prod_{x, \mu} d U_{\mu}( x) e^{ + {N \over 2\lambda} \sum_P {\rm tr} (U_P + U_P^\dagger)} \,.
\end{equation}
We work with the normalized expectation value $w(C)$, defined such that the empty loop is unity:
\begin{equation}\label{eq:Wnormalization}
\langle w(C) \rangle = {1\over N} \langle W(C) \rangle \,, \qquad w({\rm tr}(\mathds{1})) = 1 \,.
\end{equation}

\paragraph{Loops and letter representation.}

A loop can be represented by a list of ordered letters, each letter representing an edge in the lattice. 
In the general 4-dimensional lattice, we have eight types of edges 
\begin{equation}
\textrm{edges of 4-dimensional lattice}: \{\ap, \am, \bp, \bm, \cp, \cm, \dpp, \dmm \} \,,
\end{equation}
where `a,b,c,d' denote the four directions, and `-' implies the flip of the direction.
Lower-dimensional lattices are treated as subspaces;
for example, a 2D plaquette can be represented as $W[\ap,\bp,\am, \bm]$. See e.g.~\cite{Guo:2025fii} for more discussion.

\paragraph{Canonical loops.} 

The set of all possible letter strings is highly redundant due to the symmetries of the theory. 
To construct a canonical basis of operators, we must identify loops that are equivalent. 
The equivalence relation is defined by three operations:
\begin{itemize}
\item 
Cyclicity: The trace is invariant under cyclic permutations of the link variables, e.g. ${\rm tr}(ABC)={\rm tr}(BCA)$. 
\item 
Lattice symmetries: The expectation value is invariant under the hypercubic symmetry group (rotations and reflections of the lattice axes).
\item
Reversal (Conjugation): The expectation values of Wilson loops are invariant under flipping the direction (Hermitian conjugate).
\end{itemize}
We define the \emph{canonical form} of a loop as the unique representative of the equivalence class generated by these symmetries.

Distinct canonical forms correspond to distinct geometric shapes. The counting of these independent shapes is summarized in Table~\ref{tab:countingLoops}. 
As some simple examples, the 9 independent loops up to length-8 in 2D can be given explicitly as: 
\begin{equation}
\begin{split}
&\textrm{Length 4}: \  
\adjustbox{valign=c}{\begin{tikzpicture}[scale=0.5]
	\draw (0, 0)--(1, 0);
	\draw (1, 0)--(1, 1)--(0, 1)--(0, 0);
\end{tikzpicture}}
\,; \quad
\textrm{Length 6}: \  
\adjustbox{valign=c}{\begin{tikzpicture}[scale=0.5]
	\draw (0, 0)--(1, 0);
	\draw (1, 0)--(2, 0)--(2, 1)--(1, 1)--(0, 1)--(0, 0);
\end{tikzpicture}}
\,; \\
&\textrm{Length 8}: \  
\adjustbox{valign=c}{\begin{tikzpicture}[scale=0.5]
	\draw (0, 0)--(1, 0);
	\draw (1, 0)--(2, 0)--(3, 0)--(3, 1)--(2, 1)--(1, 1)--(0, 1)--(0, 0);
\end{tikzpicture}}
\,,\quad 
\adjustbox{valign=c}{\begin{tikzpicture}[scale=0.5]
	\draw (0.2, 0)--(1, 0);
	\draw (1, 0)--(1, 1);
	\draw [fill=white, draw=white] (1, 0.8) circle (0.1);
	\draw (1, 1)--(0, 1)--(0, -0.2)--(1.2, -0.2)--(1.2, 0.8)--(0.2, 0.8)--(0.2, 0);
\end{tikzpicture}}
\,,\quad
\adjustbox{valign=c}{\begin{tikzpicture}[scale=0.5]
	\draw (0, 0)--(1, 0);
	\draw (1, 0)--(1.2, 1);
	\draw [fill=white, draw=white] (1.1,0.5) circle (0.2);
	\draw (1.2, 1)--(2.2, 1)--(2.2, 0)--(1.2, 0)--(1, 1)--(0, 1)--(0, 0);
\end{tikzpicture}}
\,,\quad
\adjustbox{valign=c}{\begin{tikzpicture}[scale=0.5]
	\draw (0, 0)--(2, 0);
	\draw (2, 0)--(2, 2)--(0, 2)--(0, 0);
\end{tikzpicture}}
\,,\quad
\adjustbox{valign=c}{\begin{tikzpicture}[scale=0.5]
	\draw (0, 0)--(2, 0)--(2, 1)--(1, 1)--(1, 2)--(0, 2)--(0, 0);
\end{tikzpicture}}
\quad
\adjustbox{valign=c}{\begin{tikzpicture}[scale=0.5]
	\draw (2, 0)--(2, 1.04)--(1.04, 1.04)--(1.04, 2)--(0, 2)--(0, 0.96)--(0.96, 0.96)--(0.96, 0)--(2, 0);
\end{tikzpicture}}
\,,\quad
\adjustbox{valign=c}{\begin{tikzpicture}[scale=0.5]
	\draw (2, 0)--(2, 1)--(0, 1)--(0, 2)--(1, 2)--(1, 1.1);
	\draw (1, 0.9)--(1, 0)--(2, 0);
\end{tikzpicture}}
\,.
\end{split}
\end{equation} 
Note that since $U_\mu U^\dagger_\mu=1$, the canonical loop has no backtrack---a path that goes out and immediately returns, while self-crossing is allowed (unlike self-avoiding walks). 
Note, however, that while backtrack segments are removed from the operator definition, the variation of such segments plays a vital role in deriving the loop equations, as we will discuss below.

\begin{table}[t]
\centering
\begin{tabular}{l | r | r | r | r | r | r | r | r | r  } 
\hline
Length  	&  4 & 6 & 8 & 10 & 12 & 14 & 16 & 18  &20 \cr \hline\hline 
\# 2D   		&  1 & 1 & 7 & 15 & 95 & 465 & 3,217 & 21,762& 159,974    \cr \hline 
\# 3D   		&  0 & 2 & 11 & 117 & 1,657 & 27,012 & 488,300&9,203,186&178,996,436   \cr \hline 
\# 4D   		&  0 & 0 & 7 & 106 & 3,304 & 109,304 & {\color{black} 3,849,514}&138,614,662& /     \cr \hline 
\end{tabular} 
\caption{Number of canonical loops in various dimensions.
To avoid double-counting, the 3D counts exclude loops that are embeddable in a 2D subspace. Similarly, the 4D counts exclude loops embeddable in 2D or 3D. The symbol `/' denotes uncomputed data.
\label{tab:countingLoops}
}
\end{table}

\subsection{Loop equations}
\label{subsec:equation}

There are further dynamical constraints that relate loops with different shapes. Below, we consider the Schwinger-Dyson (SD) equations and trace relations. 

\paragraph{SD equations.}
SD equations are the quantum version of the Euler-Lagrangian equations, which are also called Makeenko-Migdal equations \cite{Makeenko:1979pb,Migdal:1983qrz} (see also \cite{Makeenko:2025bdu}).
For SD equations in lattice YM, the detailed derivations can be found in \cite{Anderson:2016rcw,Kazakov:2022xuh,Guo:2025fii}. Here we only point out some essential features that will be used below.

SD equations are derived from the variation of a Wilson link
\begin{align}
\int \mathcal{D}U \delta_{\epsilon(\mu)}[e^{-S}W^{ab}_x(\mu,C)]=0 \,. 
\end{align}
This variation has two terms:
\begin{align}
-\langle W^{ab}_x\delta_\epsilon S \rangle+\langle \delta_\epsilon W^{ab}_x \rangle=0 \,. 
\label{eq:variation of S}
\end{align}
A simple example is
\begin{equation}\label{eq:SDexample}
\begin{split}
\adjustbox{valign=c}{\begin{tikzpicture}[scale=0.5]
	\draw (0, 0)--(1, 0);
	\draw [red,midarrow] (1, 0)--(1, 1);
	\node[scale=0.8] at (1.3,0.5) {$\epsilon$};
	\draw (1, 1)--(0, 1)--(0, 0);
\end{tikzpicture}}
\,\overset{\textrm{2D}}{\longrightarrow}\quad&
\left(\,\adjustbox{valign=c}{\begin{tikzpicture}[scale=0.5]
	\draw [] (0, 0)--(1, 0);
	\draw (1, 0)--(1.2, 1);
	\draw [fill=white, draw=white] (1.1,0.5) circle (0.2);
	\draw (1.2, 1)--(2.2, 1)--(2.2, 0)--(1.2, 0)--(1, 1)--(0, 1)--(0, 0);
\end{tikzpicture}}
\,-
\adjustbox{valign=c}{\begin{tikzpicture}[scale=0.5]
	\draw [] (0, 0)--(1, 0);
	\draw (1, 0)--(2, 0)--(2, 1)--(1, 1)--(0, 1)--(0, 0);
\end{tikzpicture}}
\,+
\adjustbox{valign=c}{\begin{tikzpicture}[scale=0.5]
	\draw [] (0.2, 0)--(1, 0);
	\draw (1, 0)--(1, 1);
	\draw [fill=white, draw=white] (1, 0.8) circle (0.1);
	\draw (1, 1)--(0, 1)--(0, -0.2)--(1.2, -0.2)--(1.2, 0.8)--(0.2, 0.8)--(0.2, 0);
\end{tikzpicture}}
\,-1 \right)
-\left(\,\adjustbox{valign=c}{\begin{tikzpicture}[scale=0.5]
	\draw [midarrow] (0, 0)--(1, 0);
	\draw (1, 0)--(1, 1)--(0, 1)--(0, 0);
	\draw [midreversearrow] (1.2, 0)--(2.2, 0);
	\draw (2.2, 0)--(2.2, 1)--(1.2, 1)--(1.2, 0);
\end{tikzpicture}}
\,-
\adjustbox{valign=c}{\begin{tikzpicture}[scale=0.5]
	\draw [midarrow] (0, 0)--(1, 0);
	\draw (1, 0)--(1, 1)--(0, 1)--(0, 0);
	\draw [midarrow] (1.2, 0)--(2.2, 0);
	\draw (2.2, 0)--(2.2, 1)--(1.2, 1)--(1.2, 0);
\end{tikzpicture}}
\,+
\adjustbox{valign=c}{\begin{tikzpicture}[scale=0.5]
	\draw [midarrow] (0, 0)--(1, 0);
	\draw (1, 0)--(1, 1)--(0, 1)--(0, 0);
	\draw [midarrow] (-0.2, -0.2)--(1.2, -0.2);
	\draw (1.2, -0.2)--(1.2, 1.2)--(-0.2, 1.2)--(-0.2, -0.2);
\end{tikzpicture}}
\,-
\adjustbox{valign=c}{\begin{tikzpicture}[scale=0.5]
	\draw [midarrow] (0, 0)--(1, 0);
	\draw (1, 0)--(1, 1)--(0, 1)--(0, 0);
	\draw [midreversearrow] (-0.2, -0.2)--(1.2, -0.2);
	\draw (1.2, -0.2)--(1.2, 1.2)--(-0.2, 1.2)--(-0.2, -0.2);
\end{tikzpicture}}
\, \right)
\\
&
+ 2\lambda \left( 1 -\frac{1}{N^2}\right)\,
\adjustbox{valign=c}{\begin{tikzpicture}[scale=0.5]
	\draw [] (0, 0)--(1, 0);
	\draw (1, 0)--(1, 1)--(0, 1)--(0, 0);
\end{tikzpicture}}=0 \,,
\end{split}
\end{equation} 
where the first line is generated from $-\langle W^{ab}_x\delta_\epsilon S \rangle$, and the second line is from  $\langle \delta_\epsilon W^{ab}_x \rangle$. 

Normally we consider variations on edges belonging to Wilson loops. However, there are extra complications: certain equations can be derived by varying the edges along a backtrack path.
A simple example is
\begin{align}
	\WLbr \ \  \overset{\rm 2D}{\underset{}{\Longrightarrow}} \quad \ & \left(\, \WLb{} -\WLc{} \,\right)-\left(\, \WLdaa{midarrow}{midarrow} -\WLdaa{midarrow}{midreversearrow} \,\right) \notag \\
	& + \left(\, \WLj{} -\WLk{}\,\right) -\left( \, \WLdac{midarrow}{midarrow} -\WLdac{midarrow}{midreversearrow} \, \right) = 0 \,. 
\label{eq:exBTloopEq}
\end{align}
They are known as \emph{backtrack equations}, see \cite{Anderson:2016rcw, Kazakov:2024ool}.

\paragraph{Trace relations.}

\begin{figure}[t]
\centering
\includegraphics[width=0.55\textwidth]{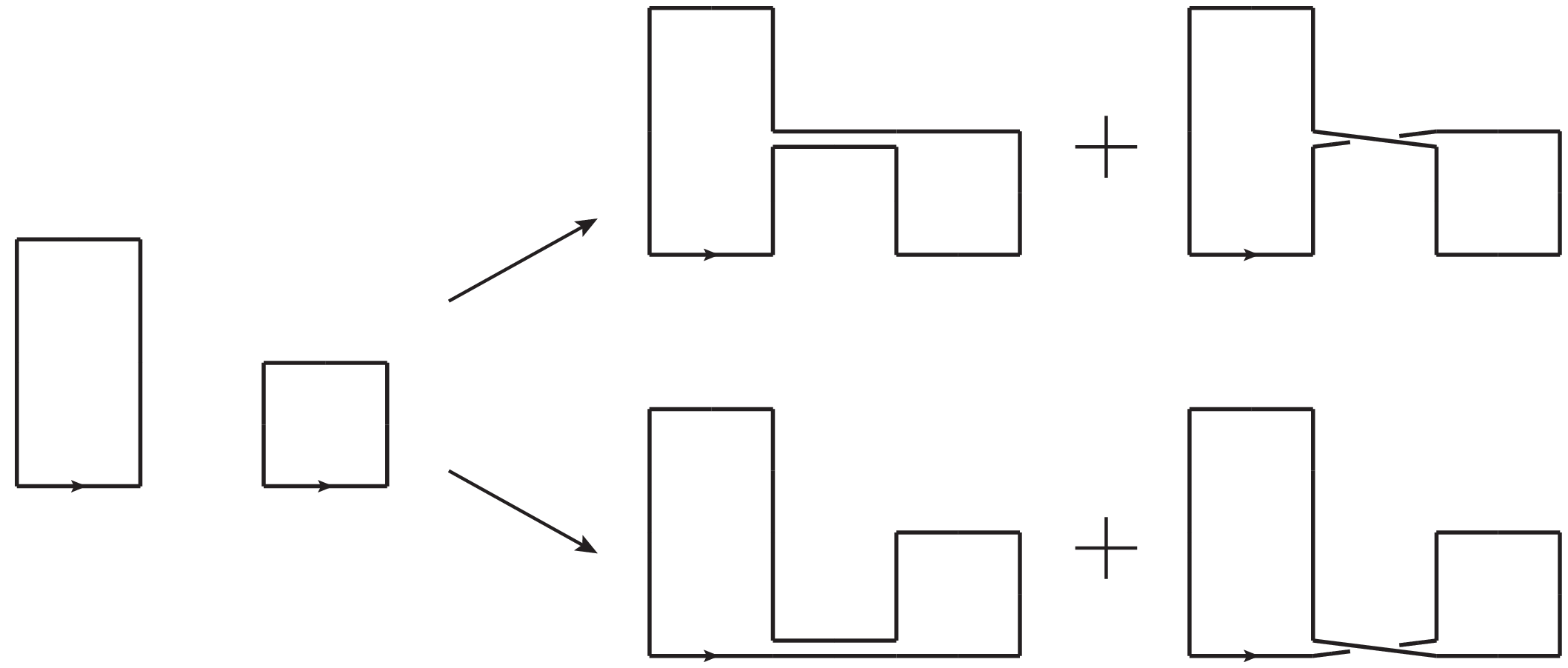}  
\caption{SU(2) trace relations.} 
\label{fig:SU2trace}  
\end{figure}

For SU(N) theories, since there are matrix identities, different shapes of loops also satisfy trace relations. Such relations are also called Mandelstam constraints \cite{Mandelstam:1978ed, Gliozzi:1979nq}.
The trace relation for SU(2) is:
\begin{align}
{\rm tr}(X){\rm tr}(Y)={\rm tr}(X Y)+{\rm tr}(X Y^\dagger) \,. 
\end{align}
With the normalization condition \eqref{eq:Wnormalization}, we have
\begin{equation}\label{eq:WccSU2app}
w(C_1) w(C_2)  = {1\over2} w(C_1\, C_2) + {1\over2} w(C_1\, \bar C_2) \,, 
\end{equation}
where both loops $C_i$ should be understood as starting from the same position $x$ as $C_{i, x}$.
A double-trace operator can be related to different pairs of single-trace operators. Consequently, the sum of these pairs of single-trace operators must be equal. One example is shown in Figure~\ref{fig:SU2trace} (see also \cite{Kazakov:2024ool}), where the two sums of single-trace operators are equal on the right-hand side:
\begin{equation}
\adjustbox{valign=c}{\begin{tikzpicture}[scale=0.5]
	\draw (0, 0)--(1, 0)--(1, 0.9)--(2, 0.9);
	\draw (2, 0.9)--(2, 0)--(3, 0)--(3, 1);
	\draw (3, 1)--(1, 1)--(1, 2)--(0, 2)--(0,0);
\end{tikzpicture}}
+
\adjustbox{valign=c}{\begin{tikzpicture}[scale=0.5]
	\draw (0, 0)--(1, 0)--(1, 0.9)--(2, 1);
	\draw [fill=white, draw=white] (1.5, 0.95) circle (0.1);
	\draw (2, 1)--(3, 1)--(3, 0)--(2, 0);
	\draw (2, 0)--(2, 0.9)--(1, 1)--(1, 2)--(0, 2)--(0,0);
\end{tikzpicture}}
=
\adjustbox{valign=c}{\begin{tikzpicture}[scale=0.5]
	\draw (0, 0)--(3, 0)--(3, 1)--(2, 1);
	\draw (2, 1)--(2, 0.1)--(1, 0.1)--(1, 2);
	\draw (1, 2)--(0, 2)--(0,0);
\end{tikzpicture}}
+
\adjustbox{valign=c}{\begin{tikzpicture}[scale=0.5]
	\draw (0, 0)--(1, 0)--(2, 0.1)--(2, 1);
	\draw [fill=white, draw=white] (1.5, 0.05) circle (0.1);
	\draw (2, 1)--(3, 1)--(3, 0)--(2, 0);
	\draw (2, 0)--(1, 0.1)--(1, 2)--(0, 2)--(0,0);
\end{tikzpicture}} \ .
\end{equation}
We will focus on SU(2) YM theory in this paper; more formulae for SU(3) can be found in \cite{Guo:2025fii}.

With trace relations, SD equations are also simplified. In SU(2), for example, \eqref{eq:SDexample} and \eqref{eq:exBTloopEq} become 
\begin{align}
& \adjustbox{valign=c}{\begin{tikzpicture}[scale=0.5]
	\draw (0, 0)--(1, 0);
	\draw [red,midarrow] (1, 0)--(1, 1);
	\node[scale=0.8] at (1.3,0.5) {$\epsilon$};
	\draw (1, 1)--(0, 1)--(0, 0);
\end{tikzpicture}}
\, \overset{\rm 2D}{\underset{\textrm{SU(2)}}{\longrightarrow}}\quad
\,\frac{3}{2}\lambda\,
\adjustbox{valign=c}{\begin{tikzpicture}[scale=0.5]
	\draw [] (0, 0)--(1, 0);
	\draw (1, 0)--(1, 1)--(0, 1)--(0, 0);
\end{tikzpicture}}
\,+
\adjustbox{valign=c}{\begin{tikzpicture}[scale=0.5]
	\draw [] (0, 0)--(1, 0);
	\draw (1, 0)--(1.2, 1);
	\draw [fill=white, draw=white] (1.1,0.5) circle (0.2);
	\draw (1.2, 1)--(2.2, 1)--(2.2, 0)--(1.2, 0)--(1, 1)--(0, 1)--(0, 0);
\end{tikzpicture}}
\,-
\adjustbox{valign=c}{\begin{tikzpicture}[scale=0.5]
	\draw [] (0, 0)--(1, 0);
	\draw (1, 0)--(2, 0)--(2, 1)--(1, 1)--(0, 1)--(0, 0);
\end{tikzpicture}}
\,+
\adjustbox{valign=c}{\begin{tikzpicture}[scale=0.5]
	\draw [] (0.2, 0)--(1, 0);
	\draw (1, 0)--(1, 1);
	\draw [fill=white, draw=white] (1, 0.8) circle (0.1);
	\draw (1, 1)--(0, 1)--(0, -0.2)--(1.2, -0.2)--(1.2, 0.8)--(0.2, 0.8)--(0.2, 0);
\end{tikzpicture}}
\,-1 = 0 \,, \\
& {\WLbr}\ \  \overset{\rm 2D}{\underset{\rm SU(2)}{\longrightarrow}} \quad \  \left(\, \WLb{} -\WLc{} \,\right) + \left(\, \WLj{} -\WLk{}\,\right) = 0 \,,
\label{eq:exBTloopEqsu2}
\end{align}
where all double-trace loops are reduced to single-trace ones.

\paragraph{Remark.} 
Before ending this section, we make two remarks. 
First, an important clarification is needed about the difference between SD equations and trace relations. 
\begin{itemize}
\item 
Trace relations are kinematic equations and apply to “off-shell” quantum loop operators $W_i$.
They are independent of the coupling constant.
\item 
SD equations are “on-shell” equations that apply to the expectation values of loops, $w_i = \langle W_i \rangle$. They generally depend on the coupling constant.
\end{itemize}
In the remainder of the paper, we will not stress their difference and will use “loop equations” to refer to both of them together. The SU(2) SD equations also encode the SU(2) trace relations directly. The figures of loops will be assumed to represent not only the loop operators but also the expectation values of the loops.

A second remark concerns the unique solvability of the 2D theory. It is well known that on a 2D lattice, the expectation values of Wilson loops factorize into products of single-plaquette averages, due to the topological simplicity of the theory \cite{Migdal:1975zg, Drouffe:1978dn, Gross:1980he, Wadia:2012fr}. 
Specifically, by adopting an axial gauge where links in a chosen direction are set to the identity, one can show that the independent observables reduce to the moments of the single plaquette, $w(n) = \langle {\rm tr}\big((U_P)^n\big) \rangle$. Using SD equations, all $w(n)$ can be further determined by the fundamental plaquette expectation value $w(1)$. 
However, we emphasize that the framework presented here does not exploit these dimension-specific simplifications, nor does it rely on gauge fixing. 
We treat the 2D case on the same footing as higher dimensions to demonstrate the general applicability of our algorithm to 3D and 4D theories.

\section{Constructing loop equations: direct ones}
\label{sec:directeqn}

In this section, we consider how to construct loop equations systematically. We address the following question: given a specific set of Wilson loops, denoted by $\mathcal{S}_0$, what is the complete set of loop equations satisfied by their expectation values $w_i = \langle W_i \rangle$ for all $W_i \in \mathcal{S}_0$?
This question is central to the bootstrap program, where $\mathcal{S}_0$ represents the finite set of operators appearing in the positivity matrices.

We will divide the loop equations into two classes:
\begin{itemize}
\item 
In the first class, every loop operator appearing in the equation (that is naturally generated as reviewed in Section~\ref{subsec:equation}) belongs strictly to the target set $\mathcal{S}_0$. We will refer to this class of equations as ``direct equations".

\item 
The second class is more subtle. In this case, the raw equations derived from variations involve extra loops that are \emph{not} contained in $\mathcal{S}_0$. However, by treating these external loops as auxiliary variables and algebraically eliminating them, one can induce effective relations purely among the loops in $\mathcal{S}_0$. Such relations will be referred to as ``indirect equations".

\end{itemize}
In this section, we focus on the direct equations, which constitute the majority of the constraints. We describe an algorithm that enables us to find \emph{all} such equations efficiently. The more elusive indirect equations will be addressed later in Section~\ref{sec:indirecteqn}.

\subsection{Plaquette-cut for SD equations}

We first consider the SD equations. 
To obtain the direct SDEs, a naive strategy would be to apply the loop variation operator to every edge of every loop in the basis $\mathcal{S}_0$. However, this brute-force approach suffers from two major drawbacks. First, it generates a large number of redundant or unrelated relations. 
Second, and more critically, it fails to systematically capture ``backtrack equations.'' These equations arise from varying edges along a backtrack path, and since there are infinitely many ways in which one can attach a backtrack path to a loop, a brute-force search is ill-defined.

To overcome these challenges, we introduce the ``plaquette-cut" strategy, which allows us to identify all relevant loop equations directly. 
This method relies on a simple geometric observation: the variation of the action, $\delta S$, effectively inserts plaquettes into the loop. Therefore, any non-trivial SD equation must involve at least one loop configuration that geometrically contains a plaquette structure.

Specifically, consider the variation term in the SDE \eqref{eq:variation of S}: 
\begin{align}
\langle W^{ab}_x\delta_\epsilon S \rangle \quad \Rightarrow \quad & \sum_{\mu\neq \pm \nu}[\langle  w_x(\mu\nu\bar{\mu}\bar{\nu}\mu C)-w_x(\nu\mu\bar{\nu}C) \rangle] + \ldots \,,
\label{eq:variation S}
\end{align}
which is generated by adding plaquettes to the variation edge and can be illustrated as
\begin{equation}
\begin{gathered} {\includegraphics[width=0.8\textwidth]{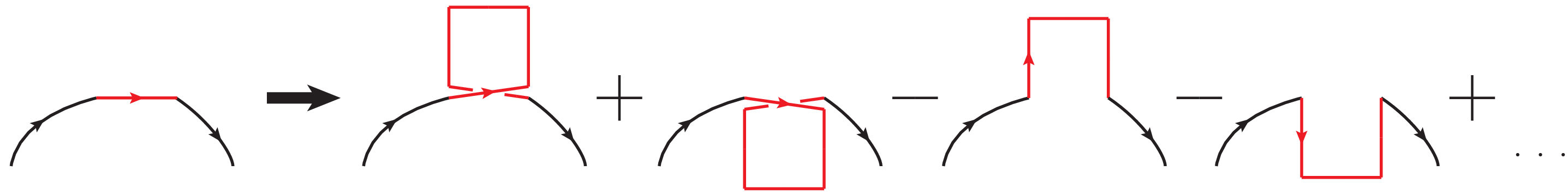} } \end{gathered} \ .
\end{equation} 
To reconstruct the SD equations, we invert this process. First, we scan the basis $\mathcal{S}_0$ for all loops containing plaquette structures. 
Next, we perform a ``cut'' on these plaquettes---excising the plaquette and determining the variation that would have produced it.
Such a procedure can generate a finite small set of candidate equations that exhaustively covers all possible direct equations.

\begin{figure}[t]
\centering
\includegraphics[width=0.5\textwidth]{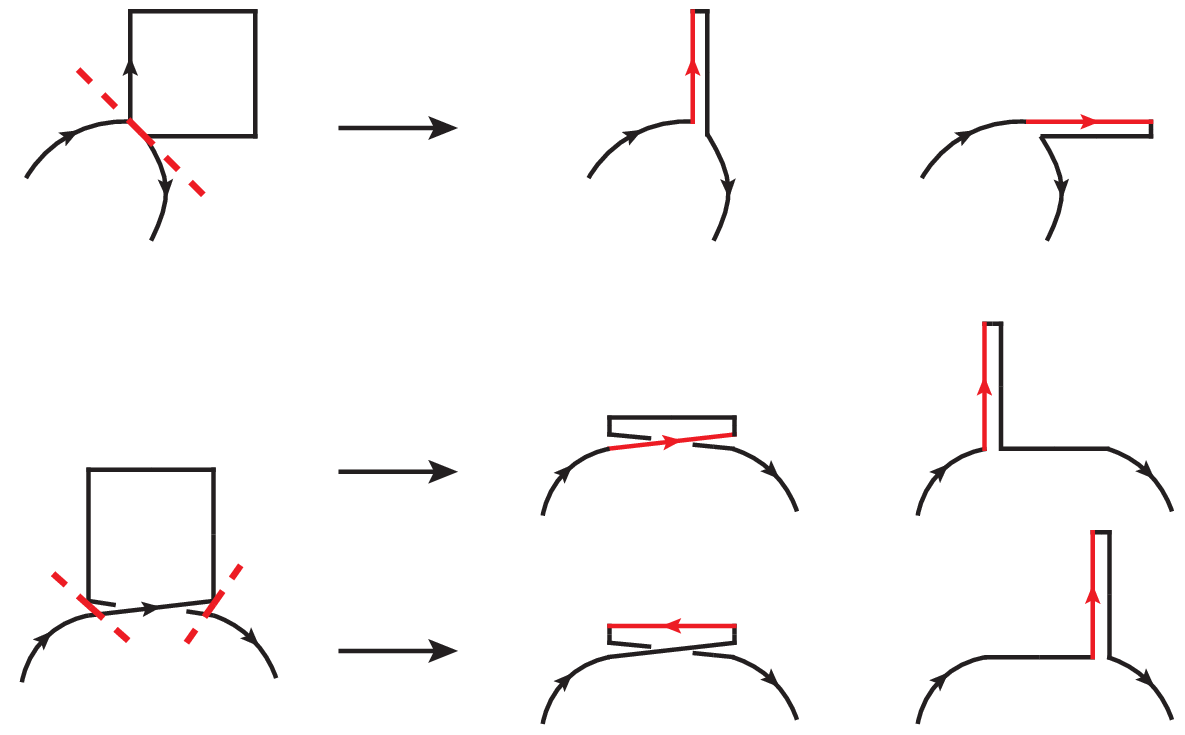}  
\caption{Plaquette-cut for direct SD equations. The red arrows represent the links for variation.} 
\label{fig:sdecut}  
\end{figure}

Consider the example in the first row of Figure~\ref{fig:sdecut}. 
We identify a plaquette structure in the loop on the left. 
Then we cut the plaquette and insert a pair of backtrack paths along the direction of the original plaquette edges. 
There are two possible ways of insertion, as shown in the right-hand figure. Finally, one generates SD equations by varying the backtrack edge (indicated in red). Such an equation necessarily contains the original loop on the LHS. If all elements appearing in the resulting equation belong to $\mathcal{S}_0$, we have successfully identified a direct equation.

Another example is shown in the second row of Figure~\ref{fig:sdecut}, where the LHS figure contains a `twisted' plaquette. Unlike the previous case, there are two ways of cutting the plaquette: at the left or right contact point. 
The cut can be described precisely in the following letter representation:
\begin{equation}
\textrm{W[$\mathcal{C}_1$,a,b,-a,-b,a,$\mathcal{C}_2$]} 
\ \overset{\textrm{cut}}{\longrightarrow} \
 \textrm{W[$\mathcal{C}_1$,\{a,b,-a,-b\},a,$\mathcal{C}_2$]} \,, \
\textrm{W[$\mathcal{C}_1$,a,\{b,-a,-b,a\},$\mathcal{C}_2$]} \,,
\label{eq:sdecut1}
\end{equation}
where $\mathcal{C}_1$ and $\mathcal{C}_2$ denote the paths preceding and following the overlapping edge ``a", and the edges enclosed in ``$\{\, \}$" are those being cut.
For each cut, we can insert a pair of backtrack paths consistent with the original plaquette's orientation (in two possible ways). So in total, there are $2\times 2$ ways of adding backtrack paths. Actually, two of them are equivalent, so this yields three independent candidate SD equations.

\begin{figure}[t]
\centering
\includegraphics[width=0.8\textwidth]{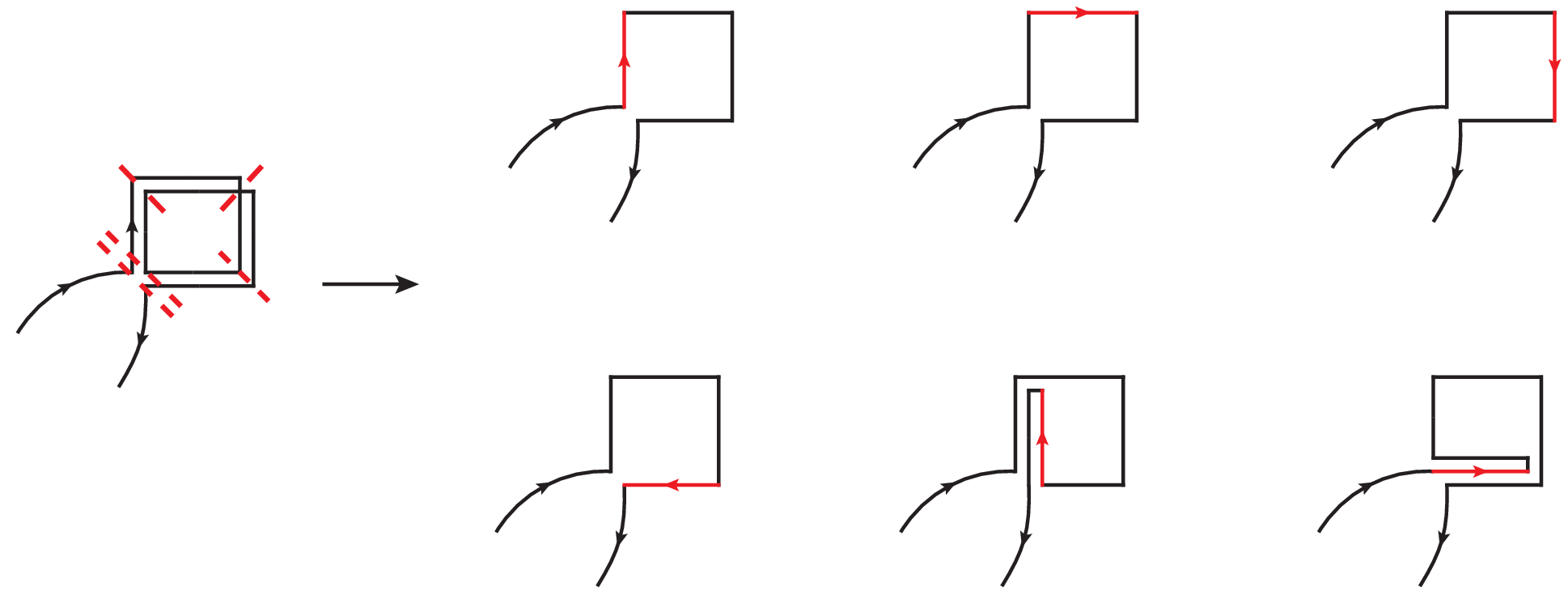}  
\caption{An example of overlapping plaquette-cut. The red arrows represent the links for variation.} 
\label{fig:sdecut2}  
\end{figure}

A more complicated example involving an overlapping plaquette is shown in Figure~\ref{fig:sdecut2}. In this case, there are five distinct ways of cutting plaquettes. In letter representation, the cuts can be given explicitly as
\begin{align}
\textrm{W[$\mathcal{C}_1$,a,b,-a,-b,a,b,-a,-b,$\mathcal{C}_2$]} 
\ \overset{\textrm{cut}}{\longrightarrow} \
& \textrm{W[$\mathcal{C}_1$,\{a,b,-a,-b\},a,b,-a,-b,$\mathcal{C}_2$]}, 
\textrm{W[$\mathcal{C}_1$,a,\{b,-a,-b,a\},b,-a,-b,$\mathcal{C}_2$]} \nonumber
\\
& \textrm{W[$\mathcal{C}_1$,a,b,\{-a,-b,a,b\},-a,-b,$\mathcal{C}_2$]}, 
\textrm{W[$\mathcal{C}_1$,a,b,-a,\{-b,a,b,-a\},-b,$\mathcal{C}_2$]}\nonumber
\\
& \textrm{W[$\mathcal{C}_1$,a,b,-a,-b,\{a,b,-a,-b\},$\mathcal{C}_2$]} .
\label{eq:sdecut2}
\end{align}
These cuts generate 10 potential backtrack insertions. After considering the equivalent ones, the total number is six, as shown in Figure~\ref{fig:sdecut2}. The variation edges are indicated by the red color, and one gets six candidate SD equations.

This ``plaquette-cut" procedure generalizes to arbitrary loop configurations. We execute this strategy for every loop in the basis $\mathcal{S}_0$. 
We can prove that this strategy gives the complete direct SD equations by proof by contradiction. 
Suppose there exists a direct equation that the algorithm fails to find. 
This equation must relate a seed loop (on which the variation is performed) to a set of resultant loops, one of which must contain the generated plaquette. Since the equation is \emph{direct}, both the seed loop and the resultant plaquette-loop must belong to $\mathcal{S}_0$.
Our algorithm iterates through every loop in $\mathcal{S}_0$, so it must be able to detect this equation. Thus, no direct equation within $\mathcal{S}_0$ can be missed.

\subsection{Subloop-cut for trace relations}

The above ``cut" strategy can be adapted for detecting trace relations. While we focus on SU(2) theory here, the approach is generalizable to higher ranks.
As reviewed in Section~\ref{subsec:equation}, SU(2) trace relations (Mandelstam constraints) relate the sum of two single-trace loops to a double-trace operator. Consequently, any single-trace loop that enters such a relation must possess a self-intersection point (a subloop structure) that allows it to be split to a double-trace operator.
To identify trace relations, we explore every $w_i \in \mathcal{S}_0$. If it contains a subloop, we `cut' it at the intersection point and obtain a double-trace operator.
Simultaneously, we construct the associated `twisted' single-trace partner and obtain a sum of two single-trace loops that potentially contribute to trace relations.
This strategy will be referred to as ``subloop-cut" strategy. 

\begin{figure}[t]
\centering
\includegraphics[width=0.8\textwidth]{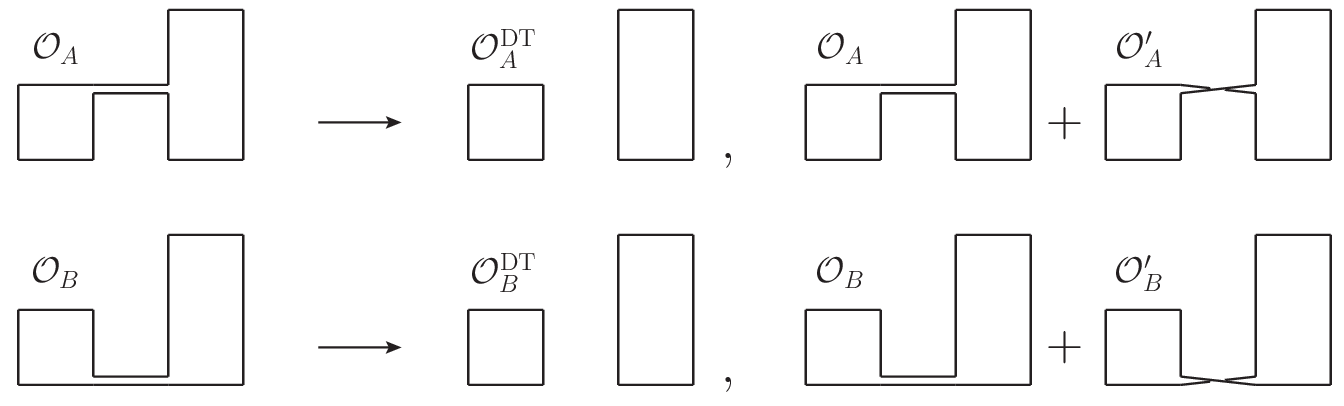}  
\caption{Subloop-cut strategy for trace relation.} 
\label{fig:subloopcut}  
\end{figure}

Consider the concrete example in Figure~\ref{fig:subloopcut}. The Wilson loop ${\cal O}_A$ on the left contains subloops, and cutting at the contact point yields the double-trace operator ${\cal O}_A^{\rm DT}$. By reversing the direction of one subloop and sewing the double-trace, we obtain the `twisted' single-trace operator ${\cal O}'_A$, which, added to the original one, gives one side of a possible trace relation. 
Now, consider a different loop ${\cal O}_B$. Its cut produces a double-trace operator ${\cal O}_B^{\rm DT}$ and a similar sum ${\cal O}_B + {\cal O}'_B$. If the two double-trace operators are identical, they will induce the following trace relation
\begin{equation}
{\cal O}_B^{\rm DT} = {\cal O}_A^{\rm DT} \quad \Rightarrow \quad {\cal O}_A + {\cal O}'_A = {\cal O}_B + {\cal O}'_B \,.
\end{equation}

Executing this subloop-cut method for all $W_i \in \mathcal{S}_0$ will generate the following set:
\begin{equation}
\big\{ {\cal O}_i^{\rm DT} \,, \ \ {\cal O}_i + {\cal O}'_i \big\} \,.
\end{equation}
For all elements in this set, the sum of two single trace operators is equal as long as the corresponding double-trace operators are the same.\footnote{Note that for SU(2) the trace is real, ${\rm tr}(U)={\rm tr}(U^\dagger)$, so the relative orientation of two subloops in the double-trace operators is irrelevant.} 
In this way, all direct trace relations within the basis are systematically generated.

\section{Growth of loops and direct equations}
\label{sec:growthofloopandeqn}

In Table~\ref{tab:2Dloopsandequations}, we show the growth of canonical Wilson loops and the direct loop equations, for loops up to length 22 in 2D lattice YM. We emphasize that here the counts of equations refer strictly to \emph{independent} equations. A few interesting patterns can be immediately seen from the table:
\begin{itemize}
\item 
The numbers of both loops and equations grow very fast, exhibiting exponential growth. We quantify this behavior below.
\item 
The SD equations (SDE) and trace relations (TrE) overlap partially, so the total number of independent equations (AllE) is less than the sum of the two individual counts.
\item
The ratio of equations to loops increases monotonically with length, suggesting that the constraints become increasingly restrictive for larger loops.

\end{itemize}

\begin{table}[t]
\centering
\begin{tabular}{c | c | c | c | c | c | c | c|c  } 
\hline
Length(2D)	&  8 & 10 & 12 & 14 & 16 & 18 &20&22  \cr \hline\hline 
\# loops  		& 9 & 24 & 119 & 584 & 3,801 & 25,563 & 185,537&1,374,414   \cr \hline 
\# SDE    		& 2  & 7 & 48 &279  &2,021  &14,665  &112,565&874,040 \cr \hline 
\# TrE   		& 0  & 0 & 6 &72  &762  &7,088  & 64,079 &561,495 \cr \hline 
\# AllE    		& 2  & 7 & 51 &316  &2,383  &17,709  & 137,775 &1,077,512 \cr \hline\hline 
{\# \textrm{AllE}}/{\# \textrm{loops}}
	  		& 0.222& 0.292 & 0.429 &0.541  &0.627 &0.693  & 0.743&0.784  \cr
\hline
\end{tabular} 
\caption{Growth of Wilson loops and the (direct) loop equations in 2D.
\label{tab:2Dloopsandequations}
}
\end{table}

\begin{figure}[t]
\centering
\subfigure[Loops and equations]{\includegraphics[width=0.48\linewidth]{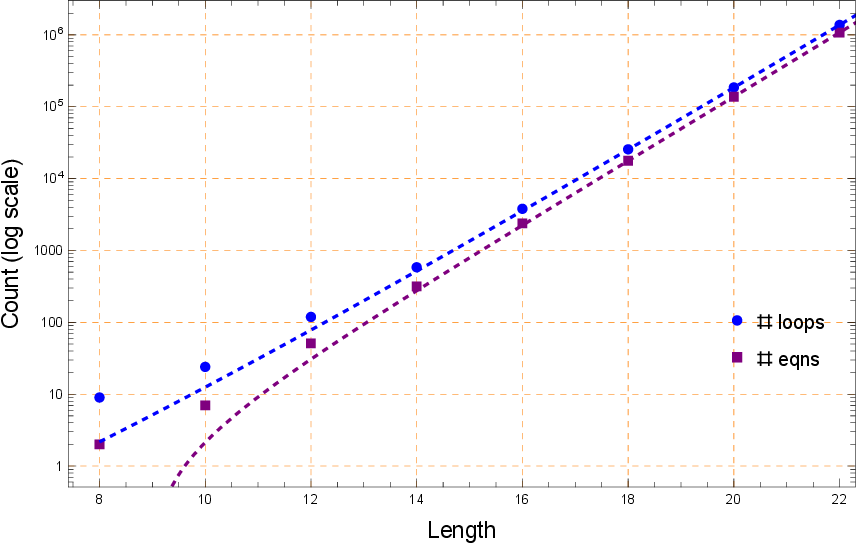}}
\qquad
\subfigure[Ratios]{\includegraphics[width=0.46\linewidth]{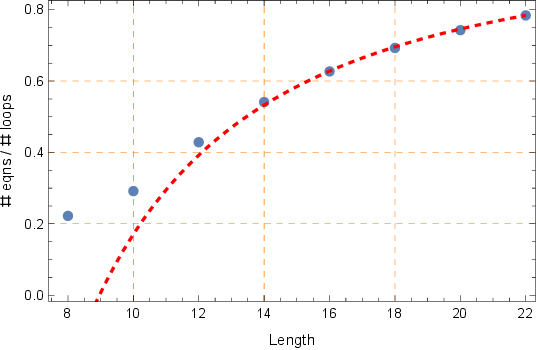}}
\caption{Growth of loops and equations.}
\label{fig:listplot}
\end{figure}

To better illustrate the trends, we plot the data in Figure~\ref{fig:listplot}.
The number of loops and equations is presented in a logarithmic plot. For large lengths, the plots become linear, which is the signature of exponential growth $N(L) \sim \mu^L$. 
We propose the following ansatz to describe the asymptotic growth of the loop basis:
\begin{equation}\label{eq:loopansatz}
N_{\rm loop}(L) \simeq A\, L^{-\alpha} \, \mu^L \,,
\end{equation}
where the parameters are defined as follows:
\begin{itemize}
\item 
$\mu^L$ is the dominant exponential growth factor. The connectivity constant $\mu=2D-1$ represents the number of available moves at each step for a non-backtracking walk on a hypercubic lattice. For the 2D lattice, $\mu=3$.\footnote{It is instructive to compare with the ``self-avoiding" walk, where $\mu\simeq 2.638$ on a 2D square lattice (see e.g.~\cite{Jensen_1999}. For 2D honeycomb lattice, the exact result $\mu=\sqrt{2+\sqrt{2}}$ is known \cite{Nienhuis:1982fx, duminilcopin2011connectiveconstanthoneycomblattice}. }
\item
$L^{-\alpha}$ is the sub-leading polynomial correction. The exponent $\alpha$ arises from two constraints. The first is the closure probability: the probability that a random walk of length $L$ returns to the origin scales as $L^{-D/2}$.\footnote{A random walk of length $L$ explores a volume $V \sim R_{\rm eff}^D$ where $R_{\rm eff} \sim \sqrt{L}$ (Gaussian statistics). The probability of ending at the origin is thus $1/V \sim L^{-D/2}$.
In contrast, for the self-avoiding case, the paths are `pushed' further and $R_{\rm eff}^{\rm SA} \sim L^{3/4}$ \cite{Flory49, Nienhuis:1982fx}, yielding $\alpha=1+3D/4$.} 
Moreover, we need to account for cyclic symmetry which enforces an $L^{-1}$ factor. Combining these two effects gives  $\alpha=1+D/2$. In a 2D lattice, we have $\alpha=2$. 
\item
$A$ is a non-universal prefactor determined by the lattice symmetries and the details of the counting, which we extract from a fit to the data.
\end{itemize}
Our asymptotic fitting function for 2D is therefore:
\begin{equation}
N_{\rm loop}^{\rm 2D} (L) \simeq A_{\rm 2D} \, L^{-2} \, 3^L \,.
\end{equation}
We can verify this ansatz by comparing the growth ratios of the data (from Table~\ref{tab:2Dloopsandequations}) against the theoretical prediction:
\begin{equation}
{N_{\rm data}^{\rm 2D}(22) \over N_{\rm data}^{\rm 2D}(20)} ={1,374,414 \over 185,537} \simeq 7.408 \,,
\qquad
{N_{\rm loop}^{\rm 2D}(22) \over N_{\rm loop}^{\rm 2D}(20)} = 3^2 \left({20 \over 22}\right)^2 \simeq 7.438 \,.
\end{equation}
The agreement is excellent.

The number of loop equations also exhibits exponential growth. Since the number of independent constraints cannot exceed the number of variables, the ratio $R(L)=N_{\rm eqn}/N_{\rm loop}$ must be bounded by unity. 
A simple ansatz for the ratio is
\begin{equation}\label{eq:ratioansatz}
R(L) \simeq R_\infty \left(1 - {c \over L^\delta} \right) \,, \qquad R_\infty \leq 1 \,,
\end{equation}
which asymptotically approaches the constant $R_\infty$ from below.
Based on the trend in Table~\ref{tab:2Dloopsandequations}, we assume that the system becomes fully constrained asymptotically, i.e.~$R_\infty=1$.\footnote{A naive fit with no restriction for $R_\infty$ would give $R_\infty>1$, which is unphysical. } 
This yields the following form for the number of equations:
\begin{equation}
N_{\rm eqn}^{\rm 2D} (L) \simeq A_{\rm 2D} \, L^{-2} \, 3^L  \left(1 - {c \over L^\delta} \right) \,.
\end{equation}
A joint fit for the loop and equation data gives
\begin{equation}
\{A_{\rm 2D} \to 0.021, \ c\to 42.134, \ \delta \to 1.706\} \,.
\end{equation}
The corresponding curves are plotted as dashed lines in Figure~\ref{fig:listplot}.
We stress that these counts include only direct equations; there are extra indirect equations which will be discussed in the next section.

Tables~\ref{tab:3Dloopsandequations} and~\ref{tab:4Dloopsandequations} present the corresponding data for 3D and 4D lattice YM.
The loop basis grows significantly faster in higher dimensions, with the generalized scaling behavior
\begin{equation}
N_{\rm loop}^{\rm 3D} (L) \simeq A_{\rm 3D} \, L^{-5/2} \, 5^L \,, \qquad
N_{\rm loop}^{\rm 4D} (L) \simeq A_{\rm 4D} \, L^{-3} \, 7^L \,.
\end{equation}
On the other hand, the ratio of equations to loops is much smaller in higher dimensions than in 2D.
This indicates that the algebraic constraints are significantly weaker in 3D and 4D, posing a greater challenge for the bootstrap method.
Additionally, we observe that in 3D and 4D, the SD equations and trace relations are largely independent of each other, in contrast to the significant overlap in the 2D case.

\begin{table}[t]
\centering
\begin{tabular}{c | c | c | c | c | c    } 
\hline
Length(3D)	&  8 & 10 & 12 & 14 & 16   \cr \hline\hline 
\# Loops & 22 & 154 & 1,906  & 29,383 & 520,900 \cr \hline 
\# SDE   & 2 & 19 & 287  & 4,913 & 93,231 \cr \hline 
\#TrE & 0 & 0 & 66  & 2,202 &58,295 \cr \hline 
\#AllE	& 2  & 19& 344 &6,785  &141,881    \cr \hline\hline
{\# \textrm{AllE}}/{\# \textrm{loops}}   	& 0.091& 0.123 & 0.180 &0.231  &0.272   \cr
\hline
\end{tabular} 
\caption{Growth of Wilson loops and the (direct) loop equations in 3D.
\label{tab:3Dloopsandequations}
}
\end{table}

\begin{table}[t]
\centering
\begin{tabular}{c | c | c | c | c |c  } 
\hline
Length(4D)	&  8 & 10 & 12 & 14 &16   \cr \hline\hline
\# Loops & 29 & 267 & 5,323  & 142,104&4,483,135  \cr \hline 
\# SDE & 2 & 20 & 443  & 11,940 &378,498 \cr \hline 
\# TrE & 0 & 0 & 110  & 5,865 &257,637 \cr \hline 
\# AllE  	& 2  & 20& 542 &17,253  &612,103    \cr \hline\hline
{\# \textrm{AllE}}/{\# \textrm{loops}}    	& 0.069& 0.075 & 0.102 &0.121&0.137     \cr
\hline
\end{tabular} 
\caption{Growth of Wilson loops and the (direct) loop equations in 4D.
\label{tab:4Dloopsandequations}
}
\end{table}

\section{Loop equations: indirect ones}
\label{sec:indirecteqn}

As mentioned at the beginning of Section~\ref{sec:directeqn}, the SDE for an operator in the subset generically produces new operators outside of the subset--- the ``operator leakage" phenomenon. Such relations can induce new relations among the variables in the subset, which are referred to as indirect equations.
In this section, we first illustrate the existence of such equations. Then we discuss strategies to detect such equations, followed by some explicit applications.

\subsection{Appearance of indirect equations}

To illustrate the mechanism, consider a toy system with two variables $\{x, y\}$. There is no direct relation between the two variables. It is possible that $x$ and $y$ may be related to other variables, say $z$, via the equations
\begin{equation}
\label{eq:xyzexample}
\left\{\begin{matrix}
x - y =  z   \\ 
x + y = 2 z 
\end{matrix}\right.
\quad \Rightarrow \quad
x = 3 y \,,
\end{equation}
which, after eliminating $z$, induces a `hidden' relation between $x$ and $y$.

We now apply this mechanism to Wilson loop variables. We use $S_{[L]}$ to denote the set of canonical loops up to length $L$. 

Consider the 2D basis truncated at length $L=12$, $S_{[12]}$. As shown in Table~\ref{tab:2Dloopsandequations}, this set comprises 119 canonical loops subject to 51 direct equations. 
Naively, this suggests a basis of $119-51=68$ independent operators. However, a crucial question arises: is this reduced basis truly irreducible, or do hidden dependencies remain?

To answer this, we enlarge the truncation to length $L=14$, $S_{[14]}$. This expanded system contains 584 loops governed by 316 direct equations, see Table~\ref{tab:2Dloopsandequations}. 
By eliminating the variables associated with length-14 loops---treating them as auxiliary intermediate degrees of freedom---we project the constraints back onto the $S_{[12]}$ subspace. This projection reveals a total of 53 constraints among the length-12 loops, two more than the 51 direct equations.\footnote{Formally, these ``indirect equations" correspond to the null space of the full equation matrix restricted to the $S_{[12]}$ subspace after eliminating those of length-14. See also the explicit example below.}
Explicitly, the two new relations are (note that the basis for these equations is not unique):
\begin{align}
\label{eq:indirecteqex1}
&\wlt-\wle+\wln-\wlq=0 \,,\\
& 1+2\lambda\left(\, \wlaa-\wlgg\right)+\lambda\left(\,\wlbb-\wlhh\right)
\nonumber\\
& \qquad\ \  -\wlpp-2\,\wlss+2\,\wltt+\wlcc-\wlii=0 \,.
\label{eq:indirecteqex2}
\end{align}
These two additional relations are \emph{indirect equations}: they are undetectable within the strict $S_{[12]}$ truncation and emerge only via the dynamical coupling to higher-length operators (here length-14 loops as \emph{intermediate} variables).

To understand the origin of the indirect equations, let us derive \eqref{eq:indirecteqex1}. Inspecting the full set of equations in the enlarged set $S_{[14]}$, we identify the following six equations:
\begin{align}
&\adjustbox{valign=c}{\begin{tikzpicture}[scale=0.5]
 	\draw (0, 0)--(0, 1)--(-1, 1)--(-1, -1);
	\draw [fill=white, draw=white] (-1,0) circle (0.2);
	\draw (-1, -1)--(-3, -1)--(-3, 0)--(-1, 0);
	\draw [red,midarrow](-1, 0)--(0, 0);
\end{tikzpicture}}
\rightarrow
\frac{3\lambda}{2}\wll-\wlk-\wle+\wlm+\wlf=0
\nonumber\\
&\adjustbox{valign=c}{\begin{tikzpicture}[scale=0.5]
	\draw (0, 0)--(0, 1)--(-1, 1);
 	\draw [red,midarrow] (-1, 1)--(-1, 0);
	\draw (-1, 0)--(-1, -1);
	\draw [fill=white, draw=white] (-1,0) circle (0.2);
	\draw (-1, -1)--(-3, -1)--(-3, 0)--(0, 0);
\end{tikzpicture}}
\rightarrow
\frac{3\lambda}{2}\,\wll-\wlk-\wln+\wlm+\wlo=0
\nonumber\\
&\adjustbox{valign=c}{\begin{tikzpicture}[scale=0.5]
 	\draw (0, 0)--(0, 1)--(-1, 1)--(-1, -0.2);
	\draw [red,midarrow] (-1, -0.2)--(0, -0.2);
	\draw (0, -0.2)--(0, -0.4)--(-1,-0.4)--(-1,-1);
	\draw [fill=white, draw=white] (-1,0.1) circle (0.2);
	\draw (-1, -1)--(-3, -1)--(-3, 0)--(0, 0);
\end{tikzpicture}}
\rightarrow
\frac{\lambda}{2}\wll+\lambda\,\wlp-\wlk+\wlt+\wlm-\wlu=0
\nonumber\\
&
\adjustbox{valign=c}{\begin{tikzpicture}[scale=0.5]
	\draw (0, 0)--(0, 1)--(-0.9, 1)--(-0.9, 0)--(-0.9, -1);
	\draw [fill=white, draw=white] (-1,0) circle (0.2);
	\draw (-0.9, -1)--(-3, -1)--(-3, 0)--(-1.3, 0);
	\draw [red,midarrow] (-1.3, 0)--(-1.3, 1);
	\draw (-1.3, 1)--(-1.1,1)--(-1.1, 0)--(0, 0);
\end{tikzpicture}}
\rightarrow
\frac{\lambda}{2}\,\wll + \lambda\,\wlp -\wlk+\wlq+\wlm-\wlr=0
\nonumber\\
&\adjustbox{valign=c}{\begin{tikzpicture}[scale=0.5]
	\draw (0,0.1)--(0,1)--(-1, 1)--(-1, 0.1)--(0, 0.1);
	\draw (0.2, 1)--(0.2, -1)--(-2, -1)--(-2, -0.1);
	\draw [fill=white, draw=white] (0.2,-0.1) circle (0.2);
	\draw (-2, -0.1)--(1.2, -0.1)--(1.2, 1)--(0.2, 1);
\end{tikzpicture}}
\rightarrow
\wln-\wlq+\wlo-\wlr=0
\nonumber\\
&\adjustbox{valign=c}{\begin{tikzpicture}[scale=0.5]
	\draw (0,-0.1)--(0,-1)--(0.9, -1)--(0.9, -0.1)--(0, -0.1);
	\draw (-0.2, 1)--(-0.2, -1)--(-2, -1)--(-2, 0.1);
	\draw [fill=white, draw=white] (-0.2,0.1) circle (0.2);
	\draw (-2, 0.1)--(0.9, 0.1)--(0.9, 1)--(-0.2, 1);
\end{tikzpicture}}
\rightarrow
\wle-\wlt+\wlf-\wlu=0 \,.
\label{indirectexample}
\end{align}
The first four are SD equations and the last two are trace relations. 
All these 6 equations contain loop variables of length 14 (shown in blue). 
By taking a linear combination of them with the coefficients
$\{1,-1,1,-1,1,-1\}$ for each equation, one can eliminate all length-14 variables and get an equation purely for loops with length $L\leq 12$, which recovers \eqref{eq:indirecteqex1}.
A similar analysis yields the second indirect equation \eqref{eq:indirecteqex2} and we give the derivation details in Appendix~\ref{app:indirecteq}.

Does the set of indirect relations saturate at this level? To investigate this, we considered even larger intermediate sets. In this case, we have considered loops up to $S_{[22]}$, but found no new constraints among $S_{[12]}$. It is therefore safe to assume that the 53 relations represent the complete set of constraints for $S_{[12]}$.

\begin{table}[t]
\centering
\begin{tabular}{l | c | c | c | c | c | c | c   } 
\hline
Length $L$ (2D)	&  8 & 10 & 12 & 14& 16 & 18 & 20    \cr \hline \hline
\# $S_{[L]}$  		& 9 & 24 & 119 & 584 & 3,801 & 25,563 & 185,537 \cr \hline\hline
\# direct eqn 	& 2  & 7& 51 & 316  & 2,383 & 17,709 & 137,775  \cr \hline 
\# Eqn from $S_{[L+2]}$   & 2  & 7& {53}&  {334}   &  {2,551} &  {19,016} &  {147,924}   \cr \hline 
\# Eqn from $S_{[L+4]}$   & 2  & 7& 53 &  {336}  &  {2,586} &  {19,325} & /   \cr \hline
\# Eqn from $S_{[L+6]}$   & 2  & 7& 53 & {336}  & {2,586} & / & /   \cr \hline
\end{tabular} 
\caption{Number of loop equations and indirect equations induced by larger loop sets in 2D.
`Eqn from $S_{[L+k]}$' refers to the total number of independent constraints projected onto the $S_{[L]}$ subspace.
\label{tab:2d indirect equation}
}
\end{table}

\begin{figure}[t]
\centering
\includegraphics[width=0.5\textwidth]{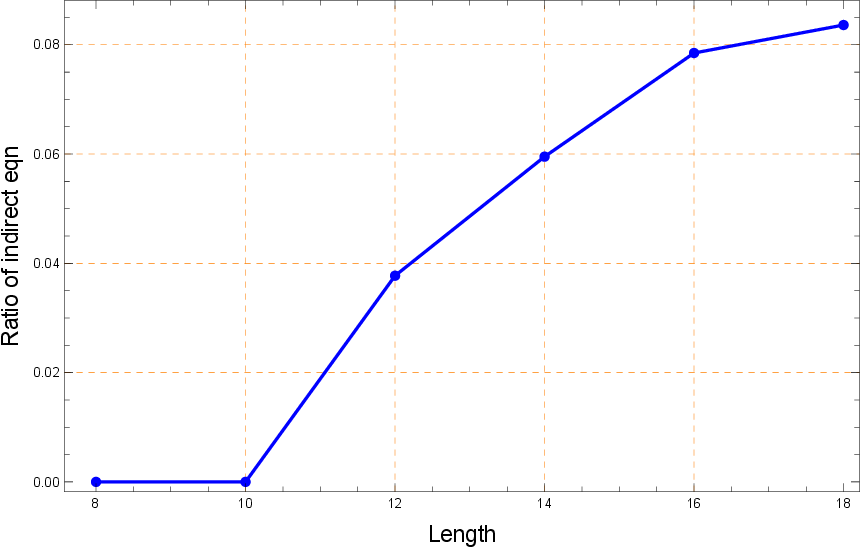}  
\caption{The growth of the ratio between indirect equations and full equations in 2D.} 
\label{fig:indirectEqnratio}  
\end{figure}

We have performed similar checks for loop sets of higher lengths, summarized in Table~\ref{tab:2d indirect equation}. 
For smaller sets ($L=8,10$), the direct equations are complete. 
However, as $L$ increases, the proportion of indirect equations grows. For example, for the initial loop set $S_{[14]}$ (316 direct equations), extending the analysis to $S_{[16]}$ reveals (334-316)=18 indirect equations among $S_{[14]}$. Extending further to $S_{[18]}$ induces two additional constraints, after which the system appears to saturate (no new constraints from $S_{[20]}$). A similar saturation pattern is observed for the $S_{[16]}$ basis.
We summarize the general behavior of indirect equations as follows:
\begin{itemize}
\item 
The number of indirect equations appears to converge quickly with respect to the size of the auxiliary space (for the cases considered). Most of the indirect equations are discovered by including loops in the immediately higher length levels.
\item
The ratio of indirect equations to total equations increases with loop length, as plotted in Figure~\ref{fig:indirectEqnratio}.
\end{itemize}
Some corresponding results for 3D and 4D loops are given in Table~\ref{tab:merged_indirect_equations}. In the 3D case, there are 3 indirect loop equation for the set $S_{[12]}$ (by enlarging up to $S_{[14]}$), the three equations can be given as
\begin{align}
&
\begin{gathered}
\includegraphics[width=0.65\textwidth]{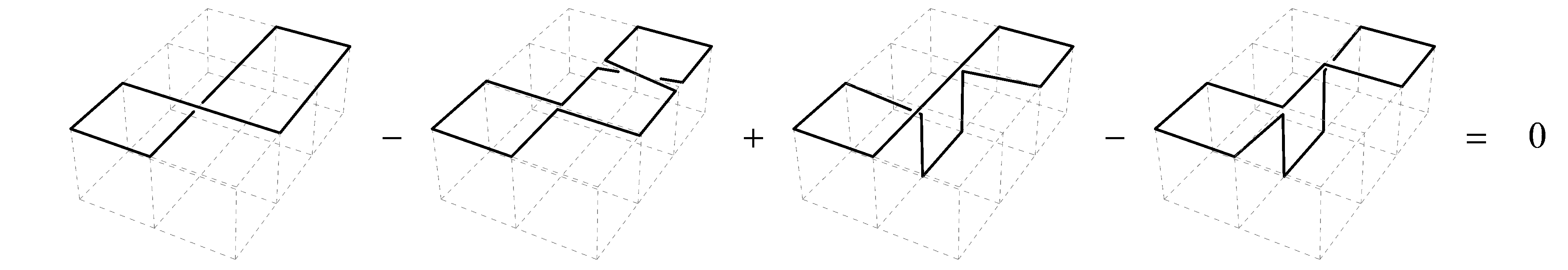}
\end{gathered}
\label{eq:indirect3D1}\\
&
\begin{gathered}
\includegraphics[width=0.8\textwidth]{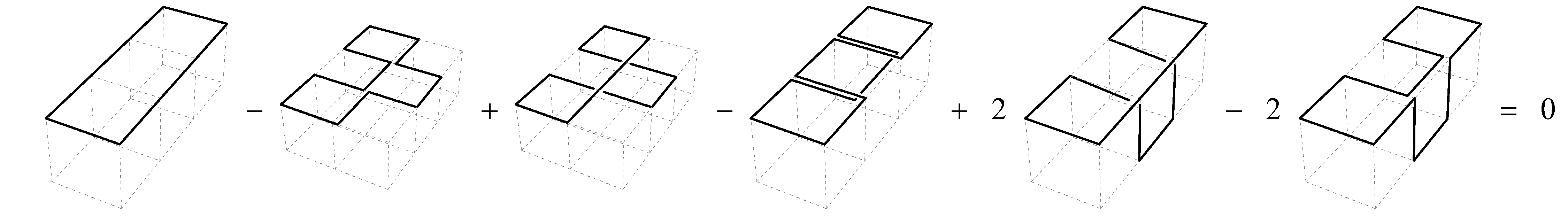}
\end{gathered}
\label{eq:indirect3D2}\\
&
\begin{gathered}
\includegraphics[width=0.92\textwidth]{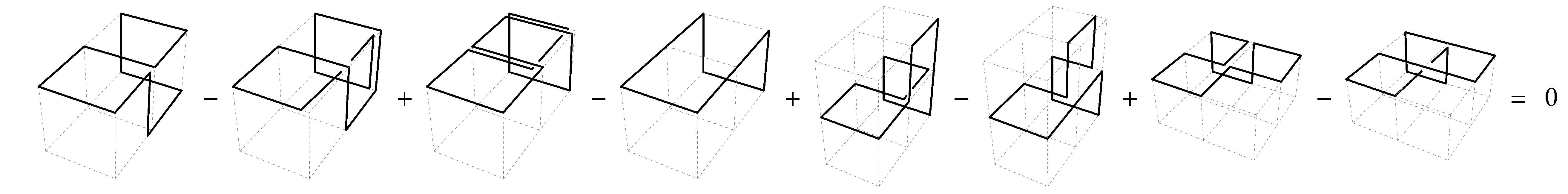}
\end{gathered}
\label{eq:indirect3D3}
\end{align}
Details of the origin of these equations are given in Appendix~\ref{app:indirecteq}.

\begin{table}[t]
\centering
\begin{tabular}{l | c|c|c|c || c|c|c|c}
\hline
& \multicolumn{4}{c||}{\textbf{3D}} & \multicolumn{4}{c}{\textbf{4D}} \\\hline\hline
Length $L$ & 8 & 10 & 12 & 14 & 8 & 10 & 12 & 14 \\ \hline \hline
\# $S_{[L]}$ & 22 & 154 & 1,906 & 29,383 & 29 & 267 & 5,323 & 142,104 \\ \hline\hline
\# direct eqn & 2 & 19 & 344 & 6,785 & 2 & 20 & 542 & 17,253 \\ \hline 
\# Eqn from $S_{[L+2]}$ & 2 & 19 &  {347} &  {6,919} & 2 & 20 &  {547} &  {17,547} \\ \hline 
\# Eqn from $S_{[L+4]}$ & 2 & 19 & 347 & / & 2 & 20 & 547 & / \\ \hline
\end{tabular}
\caption{Number of loop equations and indirect equations induced by larger loop sets in 3D and 4D.
`Eqn from $S_{[L+k]}$' refers to the equations that are projected to the $S_{[L]}$ subspace.}
\label{tab:merged_indirect_equations}
\end{table}

\subsection{Improved filtering strategies}

In practical bootstrap computations, the basis ${\cal S}_0$ is typically a specific set of loops appearing in the positivity matrices. 
They are typically a small subset of $S_{[L]}$ up to a certain length $L$.
Consequently, the brute-force approach of generalizing the basis to the full set $S_{[L+2k]}$ to find indirect equations becomes computationally inefficient.
The natural question is: can one find those indirect equations more directly? 

Since generating indirect equations requires including new loops as intermediate variables, the question can be translated as: 
how can one choose as few intermediate loops as possible to get most of the indirect equations?

A standard approach is to generate the intermediate set via the equations of motion themselves. Given an initial set $\mathcal{S}_0$, we can first generate all SDEs by variation on each edge of loops, and then collect all loops appearing in SDEs as intermediate variables. Similarly, one can also consider backtrack equations by inserting one pair of backtrack paths on loops in $\mathcal{S}_0$ and then collecting loops in the equations.

\begin{figure}[t]
\centering
\includegraphics[width=0.55\textwidth]{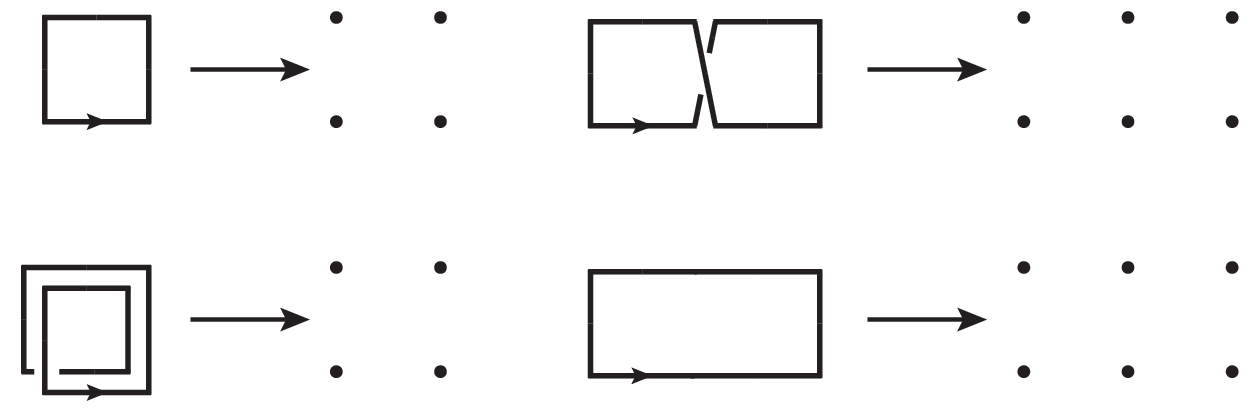}  
\caption{Vertex structure of loops. The two loops on the left share an identical set of visited vertices, as do the pair on the right.} 
\label{fig:loopstructure}  
\end{figure}

However, the set of intermediate variables generated in this way via SDEs is often still very large. To select variables which are more relevant to indirect equations, we propose a heuristic method of reducing variables. The selection is based on judging whether loops have the same ``vertex-structure" (or vertex support) as loops in $\mathcal{S}_0$. The vertex-structure is defined as below:
For a given loop variable, we record all the points traversed by the path. Overlapping points are recorded only once. Examples are given in Figure~\ref{fig:loopstructure}.  
Note that the vertex-structure respects lattice symmetry like Wilson loops, namely, rotating or reflecting the point set will give equivalent ones.

One motivation for considering ``vertex-structure" is from the empirical observation that the loops appearing in an indirect equation typically share the same vertex-structure.
Take \eqref{indirectexample} as an example, all vertex-structures of length-14 loops appear in lower-length loops.
Similar patterns are also observed in other cases, see Appendix~\ref{app:indirecteq} for another example.

In practice, one can take the following strategy. First, one finds the vertex-structure for all loops in the initial set $\mathcal{S}_0$. Next, for an enlarged set up to a certain length or via loop equations, say $\mathcal{S}_1$. Furthermore, one can `filter' this set by keeping loops that have the same vertex-structures, which gives the set $\mathcal{S}_{1,{\rm filt}}$. Finally, one generates all direct equations for $\mathcal{S}_{1,{\rm filt}}$ and project the constraints back onto $\mathcal{S}_0$.
As demonstrated below, this filtering significantly reduces the computational cost while retaining the vast majority of indirect equations.

\subsection{Applications}

We benchmark this strategy against several test cases, summarized in Table~\ref{tab:pointstructure}.
\begin{itemize}
\item 
For case I, we consider the full set of 2D loops up to length 16.  Generating intermediate variables via SDEs yields an enlarged set of 121,223 loops. With this set one obtains 2,586 equations, which are complete (according to Table~\ref{tab:2d indirect equation} using ${S}_{[20]}$).
Now we filter the loop set using the "vertex-structure" which reduces intermediate variables to 79,884. Crucially, we can still get complete indirect equations. 

\item 
Case II considers 3D loops up to length 12. In this case the filter operation significantly reduces the loop set via SDE, from  225,570 to 38,322, without missing a single indirect equation.
\end{itemize}

\begin{table}
\centering
\begin{tabular}{l|c|c|c|c|c} 
Cases &I (2D) &II (3D) &III (2D)  &IV (3D) &V (4D) \cr \hline\hline
\# loops from matrices &3,801&1,906 &8,335 &1,215 &1,011 \cr \hline 
\#  direct eqn & 2,383&344 &7,845 &461&219 \cr \hline \hline
 \# enlarged loops (SDE) &121,223 & 225,570 & 352,751  &170,611 &277,456 \cr \hline 
\# eqn & 2,586 &347  &8,062 &529&231 \cr \hline\hline
 \# enlarged loops (filter) &  {79,884}&  {38,322} &  {221,145} &  {21,937}& {24,101} \cr \hline 
\# eqn & {2,586}&   {347} & {8,045} & {525}&   {231} \cr \hline 
\end{tabular}
\caption{Indirect equations for different loop systems.}
 \label{tab:pointstructure}
\end{table} 

The above two cases confirm the efficacy of the filter strategy. Next we apply this strategy to specific bootstrap systems.
\begin{itemize}
\item
Case III is a 2D case appearing in \cite{Kazakov:2024ool} (the 2D $\Lambda=3$ level case). 
The positivity matrices contain 8,335 loops subjected to 7,845 direct equations. Enlarging the loop set via SDEs reveals (8062-7845)=217 indirect equations. With vertex filtering, the loop set reduces, yet one can still get the majority (200) indirect equations. 
\item
Cases IV and V concern some small-scale systems in 3D and 4D considered in \cite{Guo:2025fii} (path choice-4 in 3D SU(2)\footnote{See the first version of \cite{Guo:2025fii} on arXiv.} and path choice-2 in 4D SU(2), respectively). In these cases, the filter strategy reduces the loop set significantly while still getting (almost) all indirect equations.

\end{itemize}

\begin{figure}[t]
\centering
\includegraphics[width=0.5\textwidth]{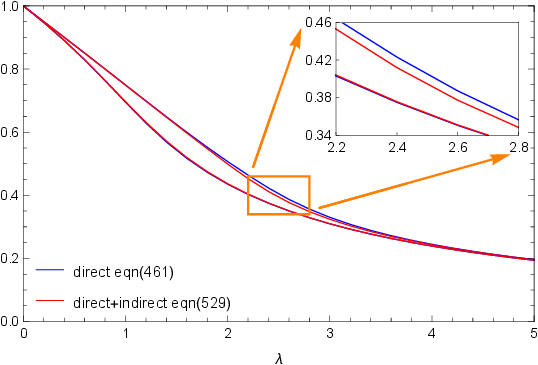}
\caption{Effect of indirect equations for the bootstrap bounds of case IV.} 
\label{fig:indirecteqnSDP}  
\end{figure}

We assess the physical impact of these indirect equations on the bootstrap bounds. For case III, we find the effect of including 200 indirect equations is negligible. 
For case IV, the improvement is notable. The system involves a $553\times 553$ positivity matrix with 1215 variables.  In Figure~\ref{fig:indirecteqnSDP}, we plot the bounds on the coupling $\lambda$ before and after adding indirect equations.
The blue lines represent the bounds using only the 461 direct equations. The red lines include indirect equations, showing a clear improvement in the upper bound from $\lambda\simeq 2$ to $\lambda \simeq 3$.

\subsection{Towards patterns of indirect equations}

Before concluding this section, we note that the set of generalized variables $\mathcal{S}_{1,{\rm filt}}$ using the above strategy remains significantly larger than the original set $\mathcal{S}_0$, typically by at least an order of magnitude.  This implies that constructing indirect equations is inherently more computationally demanding than finding direct equations. Therefore, developing more efficient strategies to derive these equations directly would be highly desirable.

We briefly outline a potential direction for future research. The main idea is that by analyzing the structure of known indirect equations, one can extract generalized construction rules.  

Consider the example of \eqref{eq:indirecteqex1}. From this relation, we can abstract a `local' structural identity:
\begin{align}\label{eq:indirecteqex1localrule}
\adjustbox{valign=c}{\begin{tikzpicture}[scale=0.5]
	\draw (0,0.1)--(0,1)--(-1, 1);
	\draw [red](-1, 1)--(-1, 0.1)--(0, -0.1);
	\draw [fill=white, draw=white] (-0.5,0) circle (0.2);
	\draw  (0, -0.1)--(0, -1)--(-1,-1);
	\draw  (-2,-0.1)--(-1, -0.1);
	\draw [red] (-1, -0.1)--(0, 0.1);
	\draw [fill=blue, draw=white] (-2,-0.1) circle (0.1);
	\draw [fill=blue, draw=white] (-1,-1) circle (0.1);
\end{tikzpicture}}
-
\adjustbox{valign=c}{\begin{tikzpicture}[scale=0.5]
	\draw (0, -1)--(0, 1)--(-1, 1);
	\draw [red](-1, 1)--(-1, 0);
	\draw (-1, 0)--(-1.2, -1);
    \draw [fill=white, draw=white] (-1.1,-0.5) circle (0.2);
	\draw (-2,0)--(-1.2, 0)--(-1, -1)--(0, -1);
	\draw [fill=blue, draw=white] (-2,0) circle (0.1);
	\draw [fill=blue, draw=white] (-1.2, -1) circle (0.1);
\end{tikzpicture}}
+
\adjustbox{valign=c}{\begin{tikzpicture}[scale=0.5]
	\draw  (0, 0)--(0, 1.2)--(-1, 1.2)--(-1, 2.2)--(1, 2.2)--(1, 1);
	\draw [fill=white, draw=white] (0,0.95) circle (0.2);
	\draw [red](1, 1)--(0,1);
	\draw (0,1)--(-1, 1);
	\draw [fill=blue, draw=white] (-1,1) circle (0.1);
	\draw [fill=blue, draw=white] (0, 0) circle (0.1);
\end{tikzpicture}}
-
\adjustbox{valign=c}{\begin{tikzpicture}[scale=0.5]
	\draw (0, 0)--(0, 1);
 	\draw (0, 1)--(-0.9, 1);
	\draw [red](-0.9, 1)--(-0.9, 0);
	\draw (-0.9, 0)--(-0.9, -1);
	\draw [fill=white, draw=white] (-0.9,0) circle (0.2);
	\draw (-2, 0)--(-2, 1)--(-1.1, 1);
    \draw [red] (-1.1, 1)--(-1.1, 0)--(0, 0);
	\draw [fill=blue, draw=white] (-2,0) circle (0.1);
	\draw [fill=blue, draw=white] (-0.9, -1) circle (0.1);
\end{tikzpicture}} 
= 0  \ ,
\end{align}
where the blue dots represent endpoints that can be connected by general paths with general lengths (provided the path does not overlap with the `variation' edges marked in red).
The validity of this identity follows from inspecting the original intermediate equations in \eqref{indirectexample}.
These equations can be generalized to loop equations associated with the following seed structures:
\begin{align}
\adjustbox{valign=c}{\begin{tikzpicture}[scale=0.5]
 	\draw (0, 0)--(0, 1)--(-1, 1)--(-1, -1);
	\draw [fill=white, draw=white] (-1,0) circle (0.2);
	\draw (-2, 0)--(-1, 0);
	\draw [red,midarrow](-1, 0)--(0, 0);
	\draw [fill=blue, draw=white] (-2,0) circle (0.1);
	\draw [fill=blue, draw=white] (-1,-1) circle (0.1);
\end{tikzpicture}}\qquad
\adjustbox{valign=c}{\begin{tikzpicture}[scale=0.5]
	\draw (0, 0)--(0, 1)--(-1, 1);
 	\draw [red,midarrow] (-1, 1)--(-1, 0);
	\draw (-1, 0)--(-1, -1);
	\draw [fill=white, draw=white] (-1,0) circle (0.2);
	\draw (-2, 0)--(0,0);
	\draw [fill=blue, draw=white] (-2,0) circle (0.1);
	\draw [fill=blue, draw=white] (-1,-1) circle (0.1);
\end{tikzpicture}}\qquad
\adjustbox{valign=c}{\begin{tikzpicture}[scale=0.5]
 	\draw (0, 0)--(0, 1)--(-1, 1)--(-1, -0.2);
	\draw [red,midarrow] (-1, -0.2)--(0, -0.2);
	\draw (0, -0.2)--(0, -0.4)--(-1,-0.4)--(-1,-1);
	\draw [fill=white, draw=white] (-1,0.1) circle (0.2);
	\draw (-2,0)--(0, 0);
	\draw [fill=blue, draw=white] (-2,0) circle (0.1);
	\draw [fill=blue, draw=white] (-1,-1) circle (0.1);
\end{tikzpicture}}\qquad
\adjustbox{valign=c}{\begin{tikzpicture}[scale=0.5]
	\draw (0, 0)--(0, 1)--(-0.9, 1)--(-0.9, 0)--(-0.9, -1);
	\draw [fill=white, draw=white] (-1,0) circle (0.2);
	\draw (-2, 0)--(-1.3, 0);
	\draw [red,midarrow] (-1.3, 0)--(-1.3, 1);
	\draw (-1.3, 1)--(-1.1,1)--(-1.1, 0)--(0, 0);
	\draw [fill=blue, draw=white] (-2,0) circle (0.1);
	\draw [fill=blue, draw=white] (-0.9,-1) circle (0.1);
\end{tikzpicture}}\qquad
\adjustbox{valign=c}{\begin{tikzpicture}[scale=0.5]
	\draw (0,0.1)--(0,1)--(-1, 1)--(-1, 0.1)--(0, 0.1);
	\draw (0.2, 1)--(0.2, -1);
	\draw [fill=white, draw=white] (0.2,-0.1) circle (0.2);
	\draw (-1, -0.1)--(1.2, -0.1)--(1.2, 1)--(0.2, 1);
	\draw [fill=blue, draw=white] (-1,-0.1) circle (0.1);
	\draw [fill=blue, draw=white] (0.2,-1) circle (0.1);
\end{tikzpicture}}\qquad
\adjustbox{valign=c}{\begin{tikzpicture}[scale=0.5]
	\draw (0,-0.1)--(0,-1)--(0.9, -1)--(0.9, -0.1)--(0, -0.1);
	\draw (-0.2, 1)--(-0.2, -1);
	\draw [fill=white, draw=white] (-0.2,0.1) circle (0.2);
	\draw (-1, 0.1)--(0.9, 0.1)--(0.9, 1)--(-0.2, 1);
	\draw [fill=blue, draw=white] (-1,0.1) circle (0.1);
	\draw [fill=blue, draw=white] (-0.2,-1) circle (0.1);
\end{tikzpicture}} \ \ \,.
\end{align}
By attaching an arbitrary path between the blue points of each structure, one recovers a system of equations analogous to \eqref{indirectexample}.\footnote{Note that the sewn path should not overlap with the variation edges marked in red, as this would change the equations.} 
Therefore, using the \eqref{eq:indirecteqex1localrule}, one can immediately generate an infinite family of indirect equations for loops of arbitrary length.

Another seed structure can be extracted by analyzing the indirect equations in 3D, as detailed in Appendix~\ref{app:indirecteq}:
\begin{equation}
\label{eq:3dstructure}
\begin{gathered}
\includegraphics[width=0.9\textwidth]{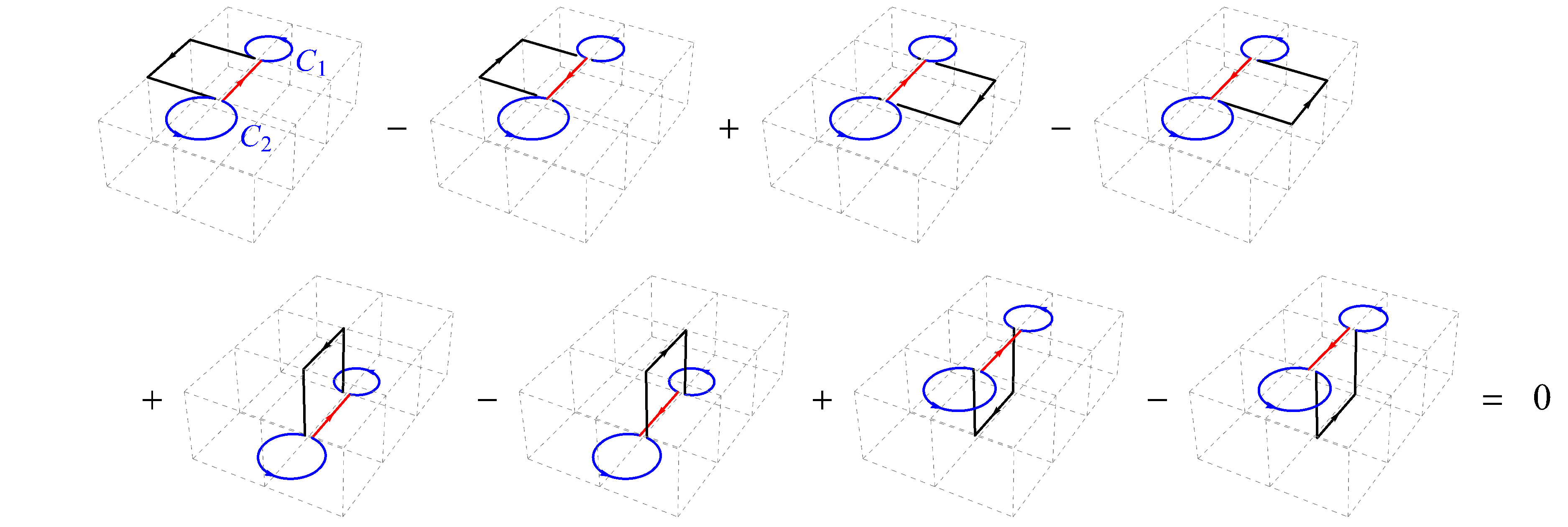}
\end{gathered}
\end{equation}
in which the blue lines, $C_1$ and $C_2$, represent two general 3D paths, provided they do not overlap with the red edge.

It would be interesting to have a systematic classification of these seed rules and have a deeper understanding of their physical origins, and we leave this problem for future study.

\section{Discussion}
\label{sec:summary}

Constructing the complete set of loop equations has been an important theoretical challenge for the lattice positivity bootstrap. 
We address this problem by introducing efficient algorithms and applying them to the concrete construction in SU(2) lattice Yang-Mills theory. 
This framework establishes a solid foundation for the bootstrap program.
While demonstrated here for SU(2), the methodology is general and should be readily extendable to higher-rank gauge groups and more complex theories.
Our study has several interesting connections and we briefly discuss them below. 

The counting of loop configurations is analogous to the study of self-avoiding walks \cite{madras2012self}. 
However, a fundamental distinction arises from the dynamical constraints imposed by gauge theory. 
While enumerating canonical loops is relatively straightforward, the construction of loop equations is a much more challenging task. 
As demonstrated in Table~\ref{tab:2Dloopsandequations}, these dynamical equations may impose severe constraints, drastically reducing the dimension of the operator basis compared to purely geometric expectations.
Following the statistical analysis of Section~\ref{sec:growthofloopandeqn}, an interesting open problem is to derive the asymptotic growth of the independent basis (including equations) analytically.

Conceptually, the problem of solving loop equations is one of basis reduction: identifying a minimal set of independent observables subject to constraints. This task finds strong resonance in other areas of theoretical physics.
In perturbative calculations, Integration-by-Parts (IBP) identities are used to reduce a vast number of Feynman integrals to a finite set of master integrals \cite{Chetyrkin:1981qh, Tkachov:1981wb}. 
Our loop equations play a role analogous to IBP identities: they are relations derived from total derivatives in the path integral that reduce the redundant set of Wilson loops to a minimal master basis.
Similarly, in continuum QFT, 
the Hilbert Series method—utilizing the Plethystic Exponential and Molien-Weyl formula—has proven to be a powerful tool for enumerating independent local operators modulo equations of motion, see e.g.~\cite{Benvenuti:2006qr, Lehman:2015via, Henning:2015daa}. 
It would be interesting to explore if a similar generating function approach can be applied for the non-local loop operators discussed here.

\acknowledgments

It is a pleasure to thank Zeyu Li, Changqi Wu, Yi-Bo Yang, and in particular, Zhiming Cai, Yuanhong Guo, and Guorui Zhu for useful discussions and suggestions at different stages of this work. 
This work is supported by the National Natural Science Foundation of China (Grants No.~12425504, 12447101, 12247103) and by the Chinese Academy of Sciences (Grant No. YSBR-101).
We also acknowledge the support of the HPC Cluster of ITP-CAS.

\appendix

\section{Plane-type loops}
\label{app:planetype}

A dimension reduction strategy is considered in \cite{Guo:2025fii}, where loops are restricted to a $(D-1)$-dimensional subspace. Such loops are referred to as plane-type loops.
In this appendix, we briefly consider the counting of loops and equations for such constrained loops.

Table~\ref{tab:plane-type 2D} and Table~\ref{tab:plane-type 3D} show the number of plane-type loops in 2D and 3D respectively. They are analogous to Table~\ref{tab:2Dloopsandequations} and Table~\ref{tab:3Dloopsandequations} for the full 2D and 3D loops. We can see that the plane-type loops  are fewer in number, while the ratio of loop equations shows small growth.

We mention that in the 4D case, all canonical loops are plane-type loops up to length 14, thus the loops and equations are the same as in Table~\ref{tab:4Dloopsandequations}. The discrepancy only starts at length 16, which will not be discussed here.

\begin{table}[t]
\centering
\begin{tabular}{c | c | c | c | c | c | c | c |c  } 
\hline
Length (2D)	&  8 & 10 & 12 & 14 & 16 & 18 &20 & 22 \cr \hline\hline
\# Loops  		& 5 & 10 & 23 & 55 & 145 & 413 & 1,280  &4,156 \cr \hline 
\# SDE   		& 1& 3 & 10 &31  &95 &303  & 1,016 &3,487 \cr
\hline
\# TrE   		& 0&0 & 0 &5  &26 &127  & 559&2,313  \cr
\hline
\# AllE    		& 1  & 3 &10 &33  &103  &337  & 1,131&3,860  \cr \hline\hline
{\# \textrm{AllE}}/{\# \textrm{loops}}  		& 0.200& 0.300 & 0.435 &0.600  &0.710 &0.816  & 0.884 &0.929 \cr
\hline
\end{tabular} 
\caption{(Direct) equations of plane-type loops in 2D.
\label{tab:plane-type 2D}
}
\end{table}
\begin{table}[t]
\centering
\begin{tabular}{c | c | c | c | c | c   } 
\hline
Length (3D)	&  8 & 10 & 12 & 14 & 16   \cr \hline\hline
\# Loops  		& 22 &154 & 1,547 & 18,438 & 234,188    \cr \hline 
\# SDE   		& 2& 19 & 236 &2,581  &30,190  \cr
\hline
\# TrE   		& 0&0 & 64 &1,894  &39,634   \cr
\hline
\# AllE    		& 2  & 3 &292 &4,249  &65,082    \cr \hline\hline
{\# \textrm{AllE}}/{\# \textrm{loops}}   		& 0.091& 0.123 & 0.189 &0.230  &0.278   \cr
\hline
\end{tabular} 
\caption{(Direct) equations of plane-type loops in 3D.
\label{tab:plane-type 3D}
}
\end{table}

The indirect equations for plane-type loops are shown in Table~\ref{tab:2d plane indirect} and Table~\ref{tab:3d plane indirect}. 
Note that $S_{[L],{\rm plane}}$ is the number of plane-type loops, which are a subset of $S_{[L]}$. 
`Eqn from $S_{[L+k]}$' refers to the equations that are projected to the $S_{[L],{\rm plane}}$ subspace.

\begin{table}[t]
\centering
\begin{tabular}{l | c | c | c | c | c | c | c |c  } 
\hline
Length $L$ (2D)	&  8 & 10 & 12 & 14& 16 & 18 & 20  & 22  \cr \hline \hline
\# $S_{[L],{\rm plane}}$ & 5 & 10 & 23 & 55 & 145 & 413 & 1,280&4,156 \cr \hline\hline
\# Direct eqn	& 1  & 3& 10 & 33  & 103 & 337 & 1,131&3,860  \cr \hline 
\# Eqn from $S_{[L]}$ 	& 1  & 3&10 & 33  & 103 & 337 & 1,131 &3,860 \cr \hline 
\# Eqn from $S_{[L+2]}$   & 1  & 3& 10& 33   & 105 & 344 & 1,158 &/  \cr \hline 
\# Eqn from $S_{[L+4]}$   & 1  & 3& 10 &33 &105 & 345 & / &/  \cr \hline
\end{tabular} 
\caption{Counting of direct and indirect equations of plane-type loops in 2D.
`Eqn from $S_{[L+k]}$' all refer to the equations that are projected to the $S_{[L],{\rm plane}}$ subspace.
\label{tab:2d plane indirect}
}
\end{table}

\begin{table}[t]
\centering
\begin{tabular}{l | c | c | c  } 
\hline
Length $L$ (3D)	&  12 & 14 & 16 \cr \hline \hline
\# $S_{[L],{\rm plane}}$ & 1,547 & 18,438 & 234,188   \cr \hline\hline
\# Direct eqn	& 292 & 4,249& 65,082    \cr \hline 
\# Eqn from $S_{[L]}$ 	& 292  & 4,271&65,670 \cr \hline 
\# Eqn from $S_{[L+2]}$   & 295  & 4,363&/  \cr \hline 
\# Eqn from $S_{[L+4]}$   & 295  & /& /  \cr \hline
\end{tabular} 
\caption{Counting of direct and indirect equations of plane-type loops in 3D.
`Eqn from $S_{[L+k]}$' all refer to the equations that are projected to the $S_{[L],{\rm plane}}$ subspace.
\label{tab:3d plane indirect}
}
\end{table}

\newpage
\section{Details of indirect equations}
\label{app:indirecteq}

In this appendix, we give more details about the origin of the indirect equations shown in the main text.

\subsection{Another 2D indirect equation}

We first consider the 2D equation \eqref{eq:indirecteqex2}.
They are related to the following intermediate equations from the enlarged set $S_{[14]}$:
\begin{align}
\adjustbox{valign=c}{\begin{tikzpicture}[scale=0.5]
	\draw [red,midarrow](-0.4, 0)--(-0.4, -0.9);
	\draw (-0.4,-0.9)--(-0.2,-0.9)--(-0.2,0)--(1,0);
	\draw [fill=white, draw=white] (0.1,0) circle (0.2);
	\draw (1,0)--(1,1)--(0,1)--(0,-1)--(-1.3,-1)--(-1.3,1)--(-0.4,1)--(-0.4,0);
\end{tikzpicture}}
&\quad \rightarrow \quad
\lambda\,\wlaa+\frac{\lambda}{2}\wlbb+\wlcc+\wldd-\wlee-\wlff=0
\nonumber\\\
\adjustbox{valign=c}{\begin{tikzpicture}[scale=0.5]
	\draw [red,midarrow](-0.1, 0)--(-0.1,-0.9);
	\draw (-0.1,-0.9)--(0.1,-0.9)--(0.1,1)--(1.2,1)--(1.2,0)--(0.4,0)--(0.4,-1.1)--(-1.1,-1.1)--(-1.1,1)--(-0.1,1)--(-0.1,0);
\end{tikzpicture}}
&\quad \rightarrow \quad
\lambda\,\wlbb+\frac{\lambda}{2}\wlaa-\wlll+\wlff+\wljj-\wloo=0
\nonumber\\\
\adjustbox{valign=c}{\begin{tikzpicture}[scale=0.5]
	\draw (-0.2,-0.8)--(-0.2,1)--(-1.2,1);
	\draw [fill=white, draw=white] (-0.1,0) circle (0.2);
	\draw (-1.2,1)--(-1.2,-1)--(0.2,-1)--(0.2,0);
	\draw [red,midarrow](0.2, 0)--(0.2,1);
	\draw (0.2,1)--(0,1)--(0,0)--(-1,0)--(-1,-0.8)--(-0.2,-0.8);
\end{tikzpicture}}
&\quad \rightarrow \quad
\lambda\,\wlgg
+\frac{\lambda}{2}\wlhh+\wlii+\wldd-\wljj-\wlkk=0
\nonumber\\\
\adjustbox{valign=c}{\begin{tikzpicture}[scale=0.5]
	\draw [red,midarrow](0.4, 0)--(0.4,1);
	\draw (0.4,1)--(0.2,1)--(0.2,-0.8)--(-0.8,-0.8)--(-0.8,0)--(0,0)--(0,1)--(-1,1)--(-1,-1)--(0.4,-1)--(0.4,0);
\end{tikzpicture}}
&\quad \rightarrow \quad
\lambda\,\wlhh+\frac{\lambda}{2}\wlgg+\wlee+\wlkk-\wlll-\wlmm=0
\nonumber\\\
\adjustbox{valign=c}{\begin{tikzpicture}[scale=0.5]
	\draw (0,0)--(0.9,0)--(0.9,1.1)--(2,1.1)--(2,2)--(0,2)--(0,0);
	\draw (1.1,0.9)--(1.1,0)--(2,0)--(2,0.9)--(1.1,0.9);
\end{tikzpicture}}
&\quad \rightarrow \quad
\wlpp+\wlqq-\wloo-\wlff=0
\nonumber\\\
\adjustbox{valign=c}{\begin{tikzpicture}[scale=0.5]
	\draw (-1,-0.8)--(-0.2,-0.8)--(-0.2,1)--(-1,1)--(-1,-0.8);
	\draw [fill=white, draw=white] (-1.1,0) circle (0.2);
	\draw [fill=white, draw=white] (-0.1,0) circle (0.2);
	\draw  (0.2, 0)--(1, 0)--(1, 1)--(0, 1)--(0, 0)--(-1.2, 0)--(-1.2, -1)--(0.2, -1)--(0.2, 0);
\end{tikzpicture}}
&\quad \rightarrow \quad
\wlss-\wltt-\wlee+\wljj=0
\nonumber\\\
\adjustbox{valign=c}{\begin{tikzpicture}[scale=0.5]
	\draw (0, 0)--(1, 0);
	\draw (1, 0)--(1, 1)--(0, 1)--(0, 0);
	\draw (-0.2, -0.2)--(1.2, -0.2);
	\draw (1.2, -0.2)--(1.2, 1.2)--(-0.2, 1.2)--(-0.2, -0.2);
\end{tikzpicture}}
&\quad \rightarrow \quad
\adjustbox{valign=c}{\begin{tikzpicture}[scale=0.5]
	\draw (0, 0)--(1, 0)--(1, 1)--(0, 1)--(0, 0);
	\draw (0, -0.2)--(0,-1)--(1,-1)--(1,-0.4)--(1.2,-0.4)--(1.2,-1.2)--(-0.2,-1.2)--(-0.2,1.2)--(1.2,1.2)--(1.2,-0.2)--(0,-0.2);
\end{tikzpicture}}
\ \rightarrow \ 
1+\wlrr-\wlkk-\wlmm=0 \,.
\label{indirectexample2}
\end{align}
%
The first four are SD equations and the last three are trace relations. 
They all contain loops of length 14 (shown in blue color). 
By taking a linear combination of them with the coefficients
$\{2,1,-2,-1,-1,-3,1\}$, one eliminates all length-14 variables and gets an equation for purely loops with length $L\leq 12$:
\begin{align}
1&+\frac{5\lambda}{2}\left(\,\wlaa-\wlgg\right)+2\lambda\left(\,\wlbb-\wlhh\right)-\wlpp+\wlrr
\nonumber\\
& - 3\,\wlss+3\,\wltt-\wlqq+2\left(\,\wlcc-\wlii\right)=0 \,.
\label{secondindirect}
\end{align}

After using two direct equations:
\begin{align}
&\adjustbox{valign=c}{\begin{tikzpicture}[scale=0.5]
	\draw (1,1)--(2,1)--(2,2)--(0,2)--(0,0)--(1,0);
    \draw [red,midarrow](1,0)--(1,1);
\end{tikzpicture}}
\rightarrow
\frac{3\lambda}{2}\wlaa+\wltt+\wlqq-\wlpp-
\adjustbox{valign=c}{\begin{tikzpicture}[scale=0.5]
	\draw (0,0)--(2,0)--(2,1)--(0,1)--(0,0);
\end{tikzpicture}}
=0 \,,
\\
&\adjustbox{valign=c}{\begin{tikzpicture}[scale=0.5]
    \draw [red,midarrow](1,0)--(1,1);
	\draw (1,1)--(0,1)--(0,0)--(1,0);
\end{tikzpicture}}
\rightarrow
\frac{3\lambda}{2}\wlgg+1+
\adjustbox{valign=c}{\begin{tikzpicture}[scale=0.5]
	\draw (0,0)--(2,0)--(2,1)--(0,1)--(0,0);
\end{tikzpicture}}
-\wlss-\wlrr=0 \,.
\label{tempeqution1}
\end{align}
One can simplify the equation to \eqref{eq:indirecteqex2}:
\begin{align}
1&+2\lambda\left(\, \wlaa-\wlgg\right)+\lambda\left(\,\wlbb-\wlhh\right)
\nonumber\\
&-\wlpp-2\,\wlss+2\,\wltt+\wlcc-\wlii=0 \,.
\end{align}

\subsection{Indirect equation in 3D}

Here we analyze in detail the three indirect equations for 3D loops given in \eqref{eq:indirect3D1}-\eqref{eq:indirect3D3}. 

The first indirect equation \eqref{eq:indirect3D1} can be constructed from following four relations:
\begin{equation}
\begin{gathered}
\begin{array}{l}
\includegraphics[width=0.8\textwidth]{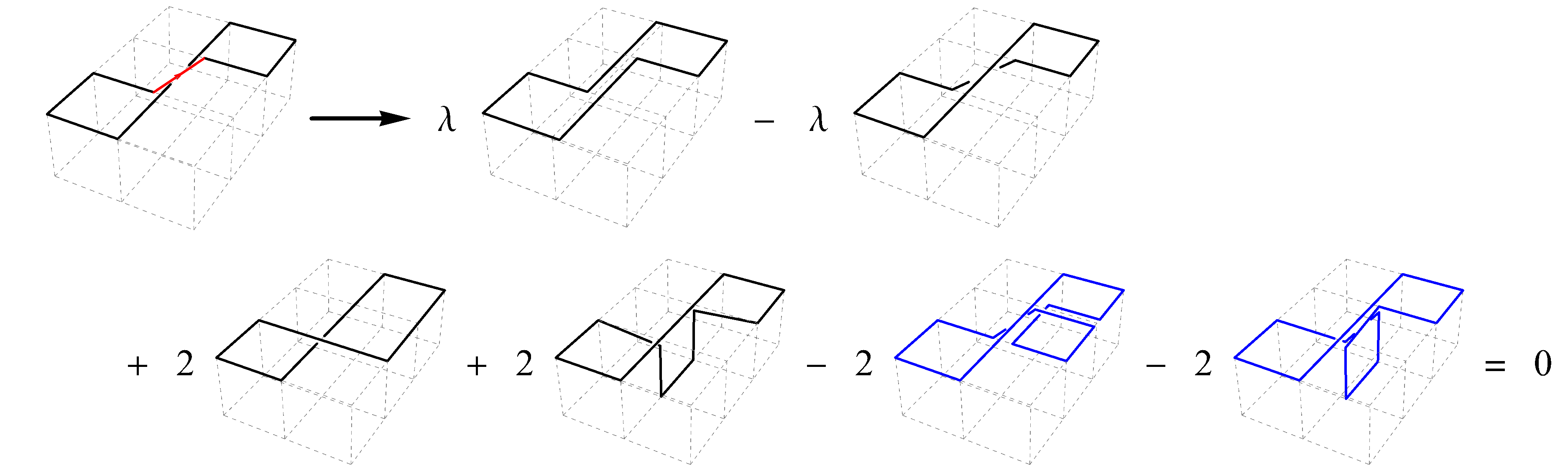}\\
\includegraphics[width=0.8\textwidth]{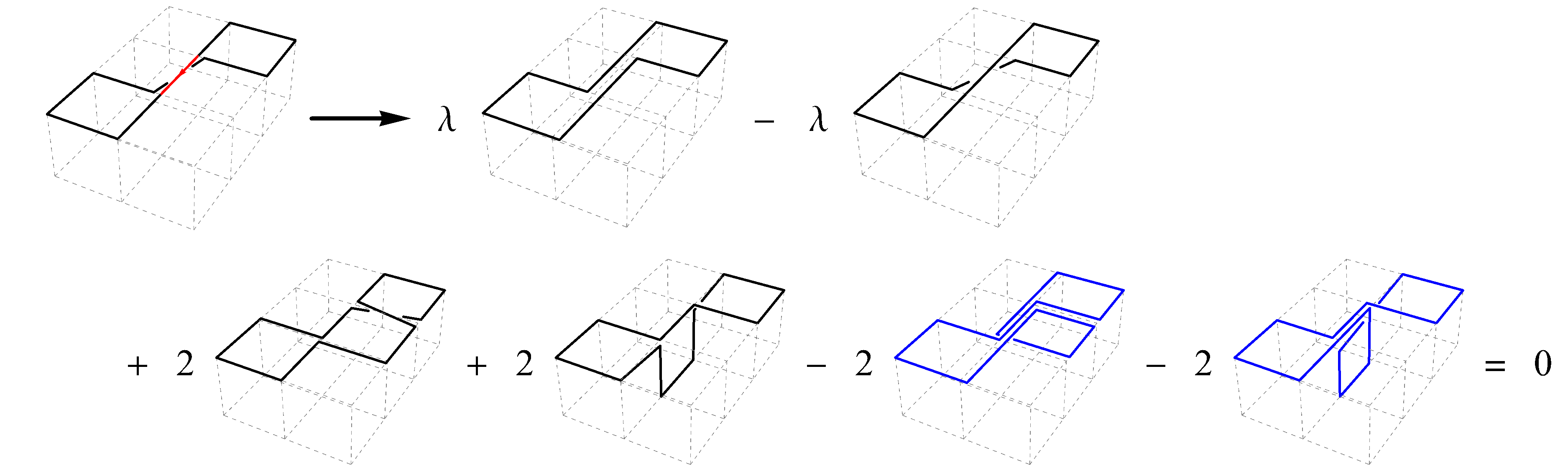}\\
\includegraphics[width=0.85\textwidth]{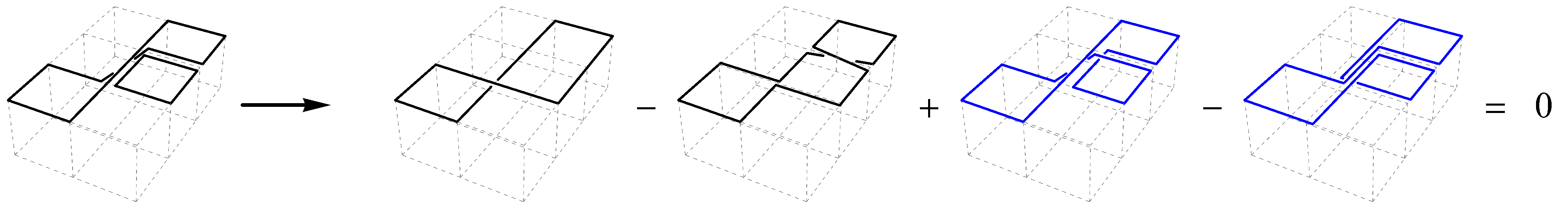}\\
\includegraphics[width=0.85\textwidth]{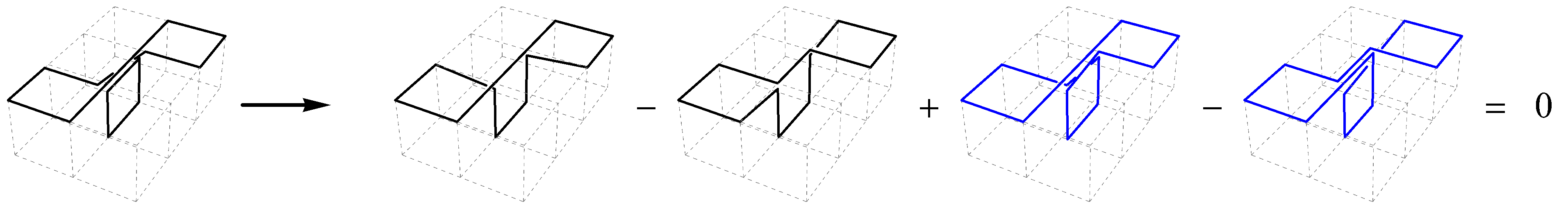}
\end{array}
\end{gathered}
\end{equation}
where the first two are SDEs, and the last two are trace relations. 
All of them contain loops of length 14, highlighted in blue color. 
By taking a linear combination with the coefficients $\{1,-1,2,2\}$, all length-14 loops are eliminated, yielding the indirect equation \eqref{eq:indirect3D1}:\begin{equation}
\begin{gathered}
\includegraphics[width=0.9\textwidth]{3dindirect1all}
\end{gathered}
\end{equation}

The second indirect equation is derived from the following five relations:
\begin{equation}
\begin{gathered}
\begin{array}{l}
\includegraphics[width=0.8\textwidth]{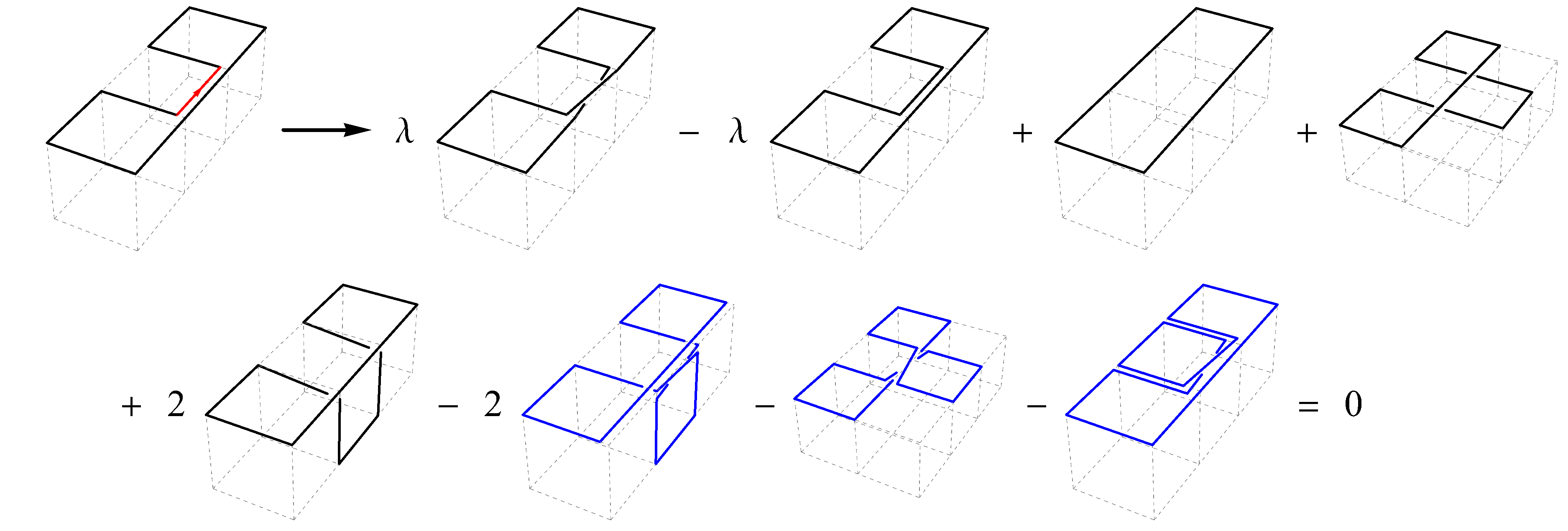}\\
\includegraphics[width=0.8\textwidth]{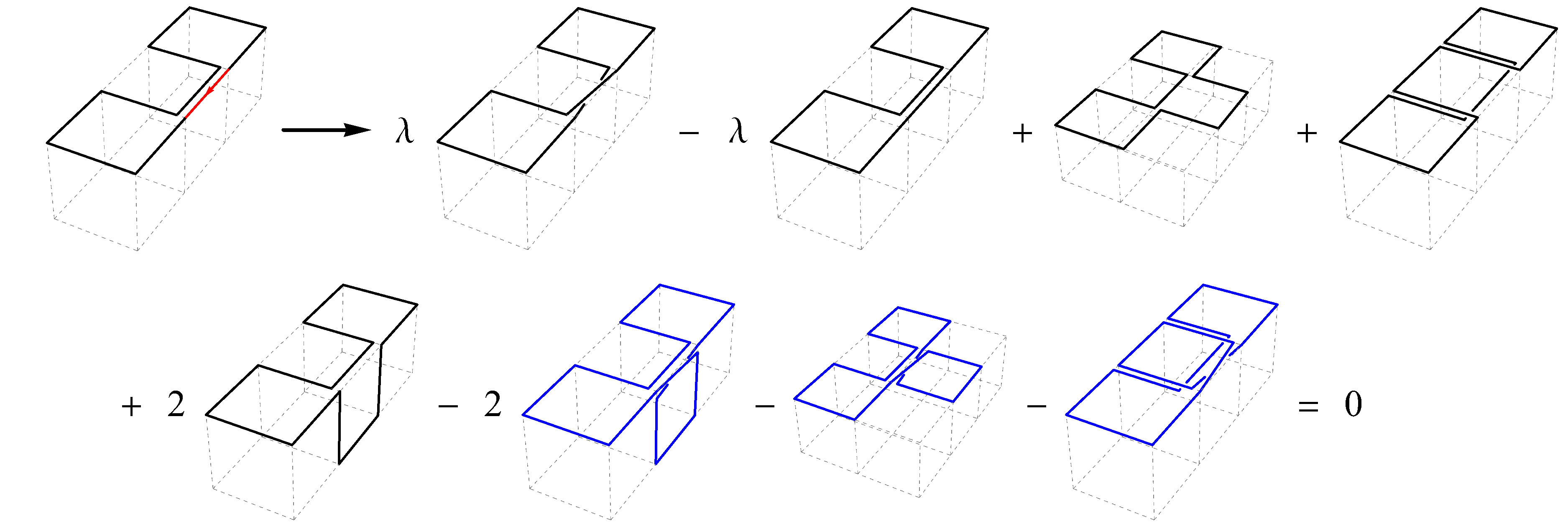}\\
\includegraphics[width=0.8\textwidth]{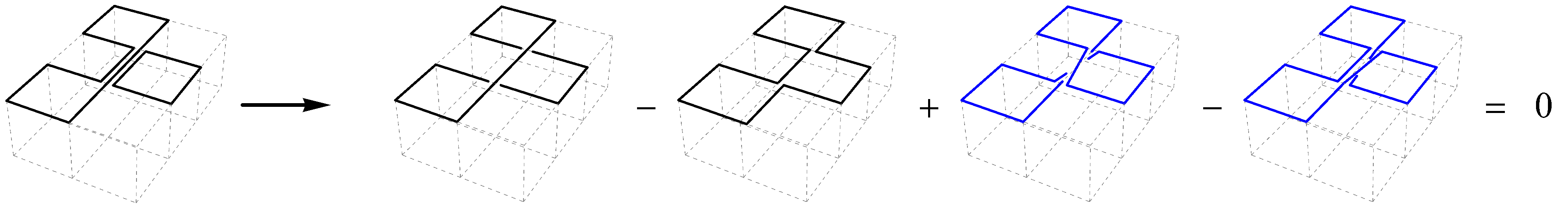}\\
\includegraphics[width=0.8\textwidth]{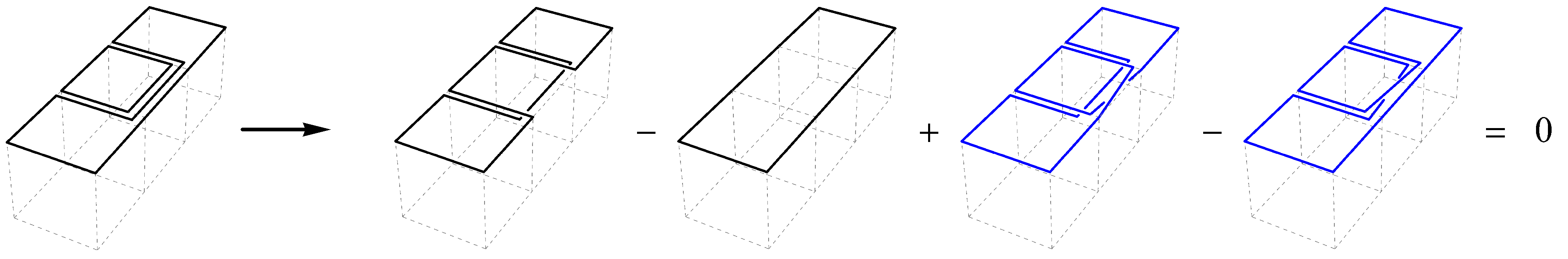}\\
\includegraphics[width=0.8\textwidth]{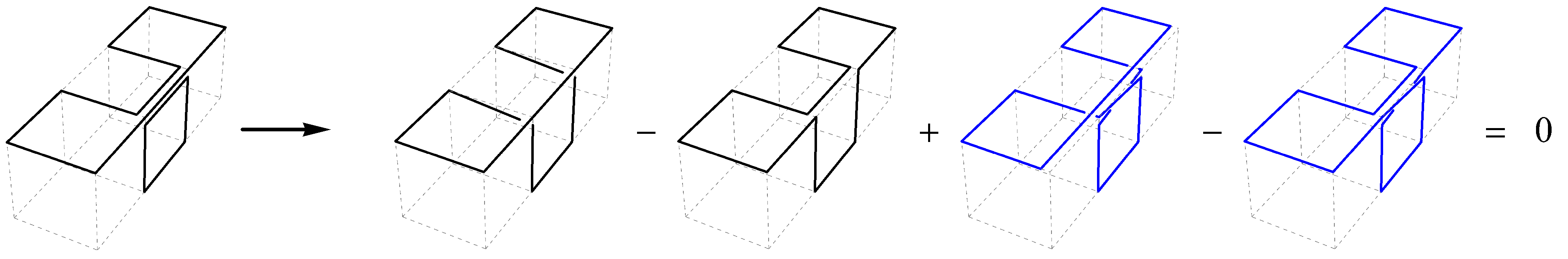}
\end{array}
\end{gathered}
\end{equation}
where the first two are SDEs, and the last three are trace relations. Taking the linear combination with the coefficients, $\{1,-1,1,-1,2\}$, all length-14 loops cancel, resulting in \eqref{eq:indirect3D2}:
\begin{equation}
\begin{gathered}
\includegraphics[width=0.9\textwidth]{3dindirect2all}
\end{gathered}
\end{equation}

The third indirect equation origins from the following six relations:
\begin{equation}
\begin{gathered}
\begin{array}{l}
\includegraphics[width=0.8\textwidth]{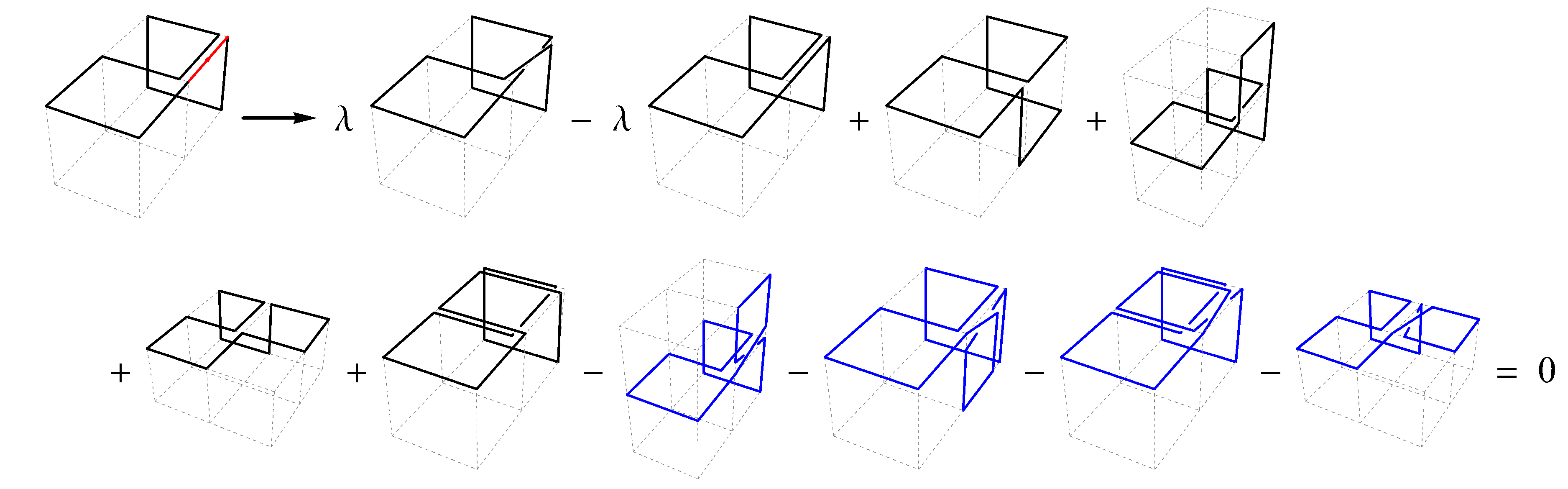}\\
\includegraphics[width=0.8\textwidth]{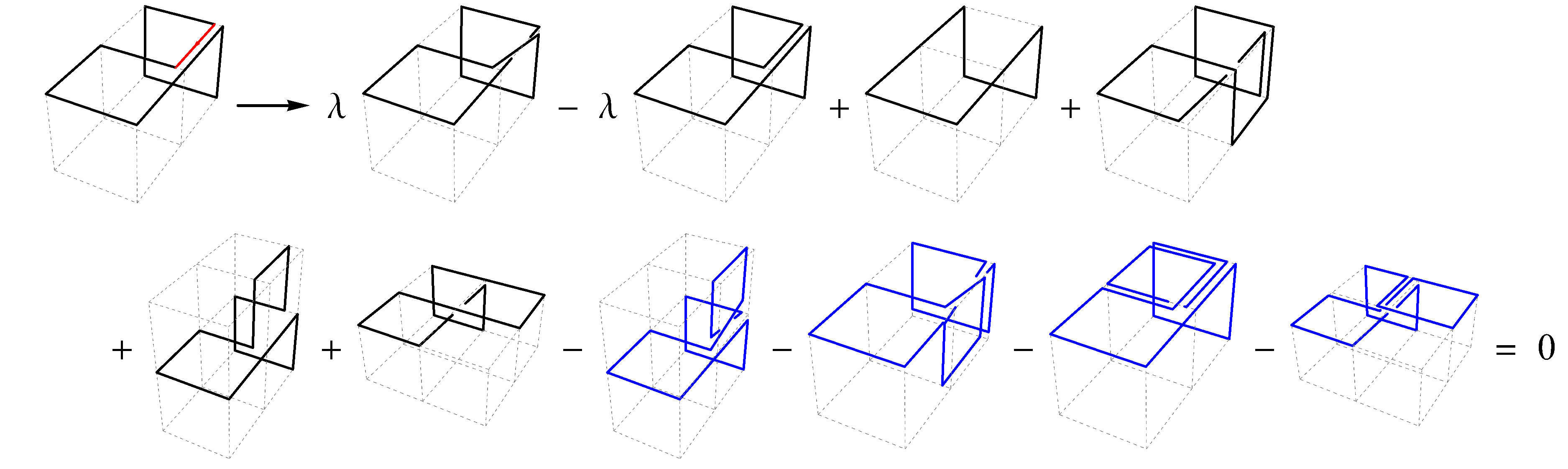}\\
\includegraphics[width=0.8\textwidth]{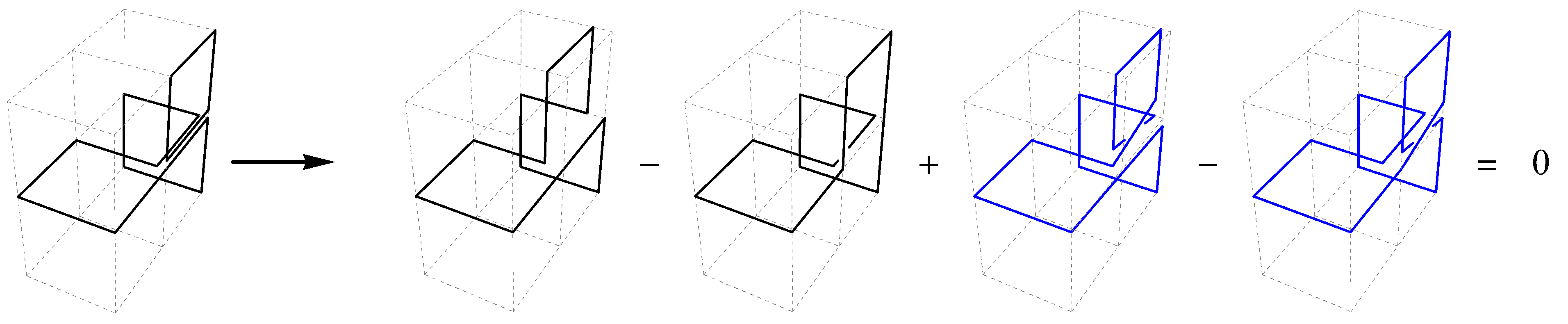}\\
\includegraphics[width=0.8\textwidth]{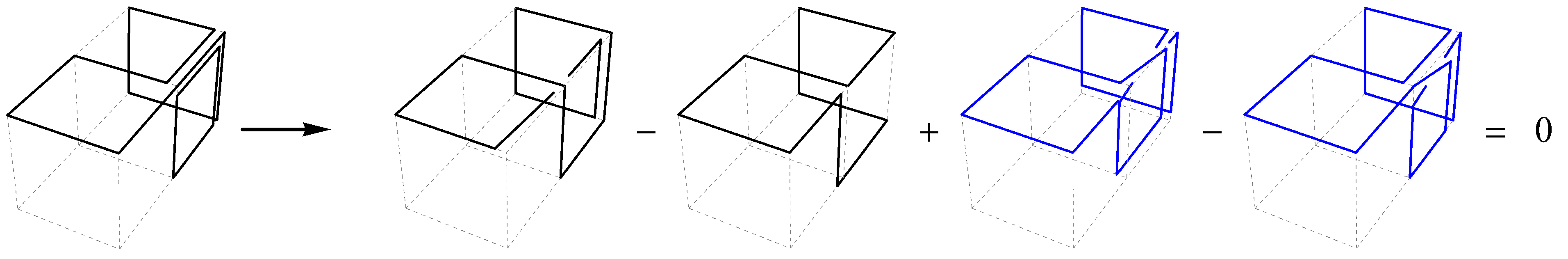}\\
\includegraphics[width=0.8\textwidth]{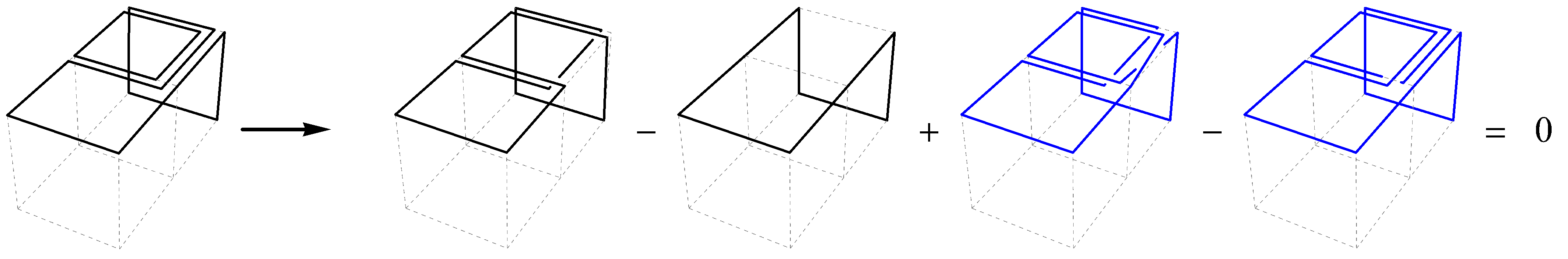}\\
\includegraphics[width=0.8\textwidth]{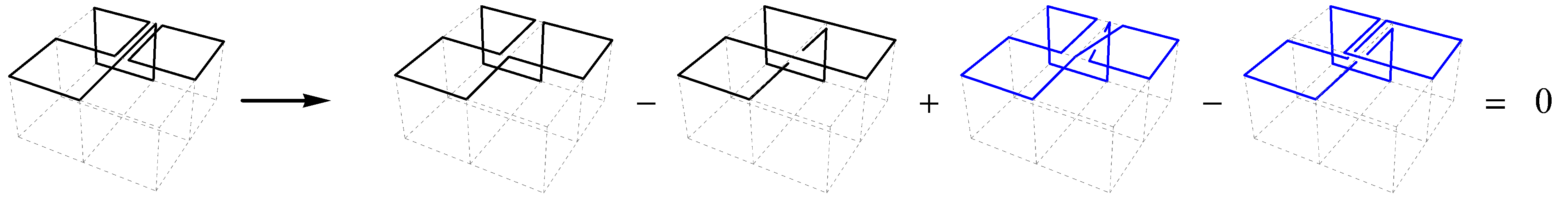}
\end{array}
\end{gathered}
\end{equation}
Combining them with coefficients, $\{1, -1, -1, -1, 1, 1\}$, we obtain \eqref{eq:indirect3D3}:
\begin{equation}
\begin{gathered}
\includegraphics[width=0.9\textwidth]{3dindirect3all}
\end{gathered}
\end{equation}

Actually, these three indirect equations share a common origin. They are all derived from the following types of equations:
\begin{equation}
\begin{gathered}
\includegraphics[width=0.9\textwidth]{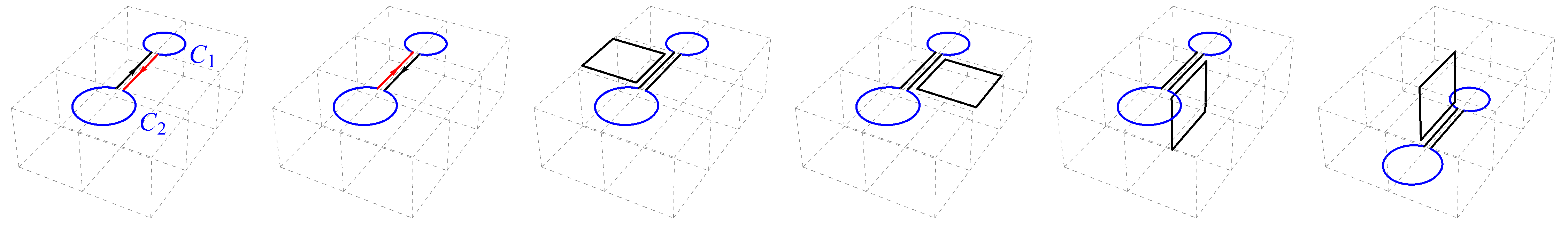}
\end{gathered}
\end{equation}
in which the blue lines represent general 3D paths, provided they do not overlap with the red variation edge.
The resulting indirect equations have the following structure:
\begin{equation}
\label{eq:3dstruc}
\begin{gathered}
\includegraphics[width=0.9\textwidth]{3dindirectstrucEqn}
\end{gathered}
\end{equation}
It is straightforward to verify that the three indirect equations above are generated by choosing $C_1$ and $C_2$ to be plaquettes. Note that \eqref{eq:indirect3D1} and \eqref{eq:indirect3D2} contain fewer terms due to the enhanced symmetry of the specific loop configurations.

The pattern \eqref{eq:3dstruc} has a direct generalization in 2D and 4D. 
\begin{itemize}
\item 
In 2D, the equation reduces to the first four terms of \eqref{eq:3dstruc}. One might expect this to generate indirect equations for $L=12$ loops, by taking $C_1$ and $C_2$ as plaquettes, analogous to the 3D case. However, it turns out that for $L=12$, these equations are not independent with the existing direct equations. Consequently, the only independent indirect equations for the set $S_{[12]}$ are those given in \eqref{eq:indirecteqex1} and \eqref{eq:indirecteqex2}. Nevertheless, this structure does generate non-trivial indirect equations for larger sets, such as $S_{[14]}$. 

\item
In 4D, due to the additional dimension, the structure equation contains four more terms (with the plaquette in \eqref{eq:3dstruc} along the fourth dimension). We have verified that such generalized equations contribute to the set of indirect equations for $L=12$ loops in 4D.
\end{itemize}

\providecommand{\href}[2]{#2}\begingroup\raggedright\endgroup


\begin{thebibliography}{10}

\bibitem{Anderson:2016rcw}
P.~D. Anderson and M.~Kruczenski, {\it {Loop Equations and bootstrap methods in
  the lattice}},  {\em Nucl. Phys. B} {\bf 921} (2017) 702--726,
  [\href{http://arxiv.org/abs/1612.08140}{{\tt arXiv:1612.08140}}].

\bibitem{Kazakov:2022xuh}
V.~Kazakov and Z.~Zheng, {\it {Bootstrap for lattice Yang-Mills theory}},  {\em
  Phys. Rev. D} {\bf 107} (2023), no.~5 L051501,
  [\href{http://arxiv.org/abs/2203.11360}{{\tt arXiv:2203.11360}}].

\bibitem{Kazakov:2024ool}
V.~Kazakov and Z.~Zheng, {\it {Bootstrap for finite N lattice Yang-Mills
  theory}},  {\em JHEP} {\bf 03} (2025) 099,
  [\href{http://arxiv.org/abs/2404.16925}{{\tt arXiv:2404.16925}}].

\bibitem{Li:2024wrd}
Z.~Li and S.~Zhou, {\it {Bootstrapping the Abelian lattice gauge theories}},
  {\em JHEP} {\bf 08} (2024) 154, [\href{http://arxiv.org/abs/2404.17071}{{\tt
  arXiv:2404.17071}}].

\bibitem{Guo:2025fii}
Y.~Guo, Z.~Li, G.~Yang, and G.~Zhu, {\it {Bootstrapping SU(3) lattice
  Yang-Mills theory}},  {\em JHEP} {\bf 12} (2025) 033,
  [\href{http://arxiv.org/abs/2502.14421}{{\tt arXiv:2502.14421}}].

\bibitem{Makeenko:1979pb}
Y.~M. Makeenko and A.~A. Migdal, {\it {Exact Equation for the Loop Average in
  Multicolor QCD}},  {\em Phys. Lett. B} {\bf 88} (1979) 135. [Erratum:
  Phys.Lett.B 89, 437 (1980)].

\bibitem{Migdal:1983qrz}
A.~A. Migdal, {\it {Loop Equations and 1/N Expansion}},  {\em Phys. Rept.} {\bf
  102} (1983) 199--290.

\bibitem{Migdal:1975zg}
A.~A. Migdal, {\it {Recursion equations in gauge field theories}},  {\em Sov.
  Phys. JETP} {\bf 42} (1975) 413--418.

\bibitem{Drouffe:1978dn}
J.~M. Drouffe and C.~Itzykson, {\it {Lattice Gauge Fields}},  {\em Phys. Rept.}
  {\bf 38} (1978) 133--175.

\bibitem{Gross:1980he}
D.~J. Gross and E.~Witten, {\it {Possible Third Order Phase Transition in the
  Large N Lattice Gauge Theory}},  {\em Phys. Rev. D} {\bf 21} (1980) 446--453.

\bibitem{Wadia:2012fr}
S.~R. Wadia, {\it {A Study of U(N) Lattice Gauge Theory in 2-dimensions}},
  \href{http://arxiv.org/abs/1212.2906}{{\tt arXiv:1212.2906}}.

\bibitem{madras2012self}
N.~Madras and G.~Slade, {\em The Self-Avoiding Walk}.
\newblock Modern Birkh{\"a}user Classics. Springer New York, 2012.

\bibitem{Wilson:1974sk}
K.~G. Wilson, {\it {Confinement of Quarks}},  {\em Phys. Rev. D} {\bf 10}
  (1974) 2445--2459.

\bibitem{Makeenko:2025bdu}
Y.~Makeenko, {\it {Notes on the Loop Equation in Loop Space}},
  \href{http://arxiv.org/abs/2508.09705}{{\tt arXiv:2508.09705}}.

\bibitem{Mandelstam:1978ed}
S.~Mandelstam, {\it {Charge - Monopole Duality and the Phases of Nonabelian
  Gauge Theories}},  {\em Phys. Rev. D} {\bf 19} (1979) 2391.

\bibitem{Gliozzi:1979nq}
F.~Gliozzi and M.~A. Virasoro, {\it {The Interaction Among Dual Strings as a
  Manifestation of the Gauge Group}},  {\em Nucl. Phys. B} {\bf 164} (1980)
  141--151.

\bibitem{Jensen_1999}
I.~Jensen and A.~J. Guttmann, {\it Self-avoiding polygons on the square
  lattice},  {\em Journal of Physics A: Mathematical and General} {\bf 32}
  (Jan., 1999) 4867–4876.

\bibitem{Nienhuis:1982fx}
B.~Nienhuis, {\it {Exact critical point and critical exponents of O(n) models
  in two-dimensions}},  {\em Phys. Rev. Lett.} {\bf 49} (1982) 1062.

\bibitem{duminilcopin2011connectiveconstanthoneycomblattice}
H.~Duminil-Copin and S.~Smirnov, {\it The connective constant of the honeycomb
  lattice equals $\sqrt{2+\sqrt2}$},
  \href{http://arxiv.org/abs/1007.0575}{{\tt arXiv:1007.0575}}.

\bibitem{Flory49}
P.~J. {Flory}, {\it {The Configuration of Real Polymer Chains}},  {\em J. Chem.
  Phys.} {\bf 17} (Mar., 1949) 303--310.

\bibitem{Chetyrkin:1981qh}
K.~Chetyrkin and F.~Tkachov, {\it {Integration by Parts: The Algorithm to
  Calculate beta Functions in 4 Loops}},  {\em Nucl.Phys.} {\bf B192} (1981)
  159--204.

\bibitem{Tkachov:1981wb}
F.~Tkachov, {\it {A Theorem on Analytical Calculability of Four Loop
  Renormalization Group Functions}},  {\em Phys.Lett.} {\bf B100} (1981)
  65--68.

\bibitem{Benvenuti:2006qr}
S.~Benvenuti, B.~Feng, A.~Hanany, and Y.-H. He, {\it {Counting BPS Operators in
  Gauge Theories: Quivers, Syzygies and Plethystics}},  {\em JHEP} {\bf 11}
  (2007) 050, [\href{http://arxiv.org/abs/hep-th/0608050}{{\tt
  hep-th/0608050}}].

\bibitem{Lehman:2015via}
L.~Lehman and A.~Martin, {\it {Hilbert Series for Constructing Lagrangians:
  expanding the phenomenologist's toolbox}},  {\em Phys. Rev. D} {\bf 91}
  (2015) 105014, [\href{http://arxiv.org/abs/1503.07537}{{\tt
  arXiv:1503.07537}}].

\bibitem{Henning:2015daa}
B.~Henning, X.~Lu, T.~Melia, and H.~Murayama, {\it {Hilbert series and operator
  bases with derivatives in effective field theories}},  {\em Commun. Math.
  Phys.} {\bf 347} (2016), no.~2 363--388,
  [\href{http://arxiv.org/abs/1507.07240}{{\tt arXiv:1507.07240}}].

\end{thebibliography}
\end{document}